\documentclass[twocolumn]{aastex631}
\usepackage{amsmath,booktabs,graphicx,hyperref,multirow}

\newcommand{\DM}{\scriptscriptstyle{{\rm DM}}}

\newcommand{\secref}[1]{Sec.~\ref{#1}}
\newcommand{\figref}[1]{Fig.~\ref{#1}}
\newcommand{\appref}[1]{Appx.~\ref{#1}}
\newcommand{\tabref}[1]{Tab.~\ref{#1}}
\renewcommand{\eqref}[1]{Eq.~(\ref{#1})}

\begin{document}

\title{Effects of Dark Matter Self Interactions on Sagittarius and Its Stream}

\author{Connor Hainje}
\affiliation{Center for Cosmology and Particle Physics, Department of Physics, New York University, New York, NY 10003, USA}
\email{connor.hainje@nyu.edu}

\author{Oren Slone}
\affiliation{Raymond and Beverly Sackler School of Physics and Astronomy, Tel Aviv University, Tel-Aviv 69978, Israel}

\author{Mariangela Lisanti}
\affiliation{Department of Physics, Princeton University, Princeton, NJ 08544, USA}
\affiliation{Center for Computational Astrophysics, Flatiron Institute, New York, NY 10010, USA}

\author{Denis Erkal}
\affiliation{School of Mathematics and Physics, University of Surrey, Guildford, GU2 7XH, Surrey, UK}

\begin{abstract}
This work explores how assumptions regarding the particle-physics nature of dark matter can alter the evolution of the Sagittarius~(Sgr) dwarf spheroidal galaxy and its expansive stellar stream.  We run a large suite of $N$-body simulations to model the infall of a Sgr-like dwarf, exploring how the presence of dark matter self interactions impacts its evolution. For a scattering cross section of $\sigma/m_\chi = 30~\text{cm}^2\text{/g}$ (at orbital velocity scales), these interactions result in significantly less stellar mass and little to no dark matter bound to the progenitor at the present day.  To isolate the cause of this mass loss, we introduce a novel technique for controlling which pairs of dark matter simulation particles can interact.  This enables us to identify ram-pressure evaporation---the scattering of satellite and host dark matter particles---as the primary source of the enhanced mass loss. The rapid disintegration of the Sgr progenitor when self interactions are allowed alters some key properties of the resulting stellar stream, most dramatically suppressing the presence of a ``spur’’ on the apocenter of the trailing stream arm that correlates with the mass of the satellite at last pericenter. We demonstrate how the effects on the Sgr system scale with the particular choice of self-interaction cross section, which affects the degree of ram-pressure evaporation. These findings generalize beyond the Sgr system, underscoring that dwarf stellar streams and dwarf galaxies with close passages may serve as sensitive probes for dark matter self interactions. Animations and three-dimensional visualizations of our simulations are available at \url{https://connorhainje.com/sgr-sidm-viz}.
\end{abstract}

\section{Introduction}
\label{sec:introduction}

Although the Lambda Cold Dark Matter~($\Lambda$CDM) paradigm is remarkably well-tested on large scales~\citep[e.g.,][]{planck}, its viability on small scales remains uncertain~\citep[e.g.,][]{Sales:2022ich}. 
A defining characteristic of CDM models is that dark matter interacts only gravitationally.  However, the possibility that dark matter interacts with itself through a new force~\citep{sidm_spergelsteinhardt} remains a well-motivated alternative that naturally arises in many theories---see, e.g.,~\citet{tulin_yu} for a review. Such scenarios, referred to as Self-Interacting Dark Matter~(SIDM), leave the large-scale dark matter distribution essentially unaffected, but result in distinctive imprints on sub-galactic scales~\citep{smallscales}. 

In this work, we focus specifically on SIDM implications for one of the most well-studied substructures in the Milky Way: the Sagittarius~(Sgr) dwarf galaxy~\citep{sgr, 10.1093/mnras/277.3.781} and its accompanying stellar stream~\citep{sgr_2MASS}. Located $\sim 24$~kpc from the Sun, Sgr is one of the closest satellites to us.  It has been significantly disrupted over the course of its orbital evolution and is expected to dissolve completely within the next Gyr~\citep{sgr_vasiliev}, assuming CDM. The degree of tidal disruption has resulted in an extensive stellar stream that wraps around the Galaxy, observed to distances of  $\sim 100$~kpc~\citep{belokurov+14}. The stream has been the subject of a strong mapping effort over the years~\citep[e.g.,][]{sgr_2MASS, sgr_SDSS, sgr_koposov+12, belokurov+14, hernitschek+17}, recently accelerated with data from \emph{Gaia} and other observatories~\citep{sgr_streamfinder, 2020A&A...638A.104R, 2020A&A...635L...3A,Penarrubia_Petersen_2021,2022A&A...666A..64R, 2023ApJ...946...66L, 2024ApJ...963...95C}. 

The Sgr stream has several notable features. It has visible leading and trailing stream arms, as first detected in \citet{sgr_2MASS}. The apocenters of the two stream arms differ substantially, with the leading arm apocenter at a Galactocentric distance of $\sim 50 \ \text{kpc}$ and the trailing arm apocenter at $\sim 100 \ \text{kpc}$ \citep{belokurov+14,hernitschek+17}. The arms wrap entirely around the Milky Way, with stars observed in at least two wraps \citep{2020A&A...635L...3A} and some models predicting another \citep[e.g.,][]{dierickx}. Both stream arms are observed to have ``bifurcated'' \citep{sgr_SDSS,sgr_koposov+12}; the origin of this bifurcation is not understood, but one hypothesis is that it results from the rotation of a disk in the initial Sgr progenitor \citep[e.g.,][]{penarrubia+2010,Oria+2022}. A more recently discovered feature of the Sgr stream is a ``spur'' on the apocenter of the trailing arm \citep{spur_sesar+17, 2019MNRAS.490.5757S, Bayer+2025}, with stars extending past the apocenter to a heliocentric distance of $\sim 130\ \text{kpc}$. Spurs like this are predicted by dynamical models such as \citet{gibbons+2014,dierickx,fardal+19}.

The degree to which a satellite galaxy like Sgr is disrupted along its orbit depends, among other factors, on the strength of dark matter self interactions, which affects mass loss in several key ways. The first is through the satellite's internal density distribution, which is altered by efficient heat flow that is generated by self interactions and is thus not present in CDM. In this picture, heat initially flows towards the center of the galaxy and rapidly thermalizes the dark matter, resulting in a centrally-cored density profile. If enough time passes, however, heat begins to flow out of the central core and into the colder outer regions of the halo. Because a gravitationally bound halo has negative specific heat, the result is gravothermal core collapse, which causes a rapid increase in the central density of the halo~\citep{Balberg:2002ue}. The timescale for core collapse can be drastically decreased via tidal stripping and can thus be relevant for the evolution of Milky Way satellites~\citep{Nishikawa:2019lsc}.  The concentration of a satellite's halo, as set through such processes in SIDM, can affect how rapidly it disrupts~\citep{slone+23} as well as the properties of its tidal debris~\citep{10.1093/mnrasl/slv012}.

The second key consequence of SIDM is that it can induce scattering between dark matter in the satellite and the host halo. These scatterings typically have a negligible effect on the satellite's orbit since momentum transfer is extremely small, due to the large hierarchy between the satellite's virial velocity and its orbital velocity~\citep{sidm_rampressure,slone+23}. Instead, most scattering events of this type result in dissolution of the satellite's dark matter halo, referred to as ram-pressure evaporation. For large-enough scattering cross sections at the relevant velocity scale, this effect can efficiently remove dark matter from all regions of the satellite and even turn off the process of core collapse~\citep{Zeng:2021ldo,slone+23}.

Demonstrating how self interactions affect Sgr requires detailed numerical modeling of the satellite's evolution in the Milky Way.  In this work, we generalize state-of-the-art $N$-body techniques developed to model Sgr in a CDM framework~\citep[e.g.,][]{sgr_law2010, sgr_purcell, gibbons+2014, gibbons+2017, dierickx, sgr_laporte, fardal+19, tango} to the case of SIDM.  In particular, we initialize the Sgr progenitor in the outskirts of an isolated Milky Way--like halo and evolve the system, which includes both dark matter and star particles, forward in time.  The initial conditions of the progenitor are optimized to reproduce the present-day location and velocity of the dwarf remnant.  Because the host halo is live, the orbital evolution includes effects like dynamical friction and ram-pressure evaporation.  

Given the complexity of reproducing the data for both the Sgr dwarf and stream---which no study to date has succeeded in doing---this work focuses on the qualitative effects that variations to dark matter modeling   induce on the system.  Because the Sgr satellite has survived two full pericentric passages remarkably close to the Galactic center~\citep[e.g.,][]{fardal+19}, it has passed through orbital regimes where the effects of ram-pressure evaporation should be most pronounced. Indeed, we find that for the self-scattering cross sections considered here, the Sgr progenitor experiences significantly more mass loss in SIDM than in CDM---to such an extent that it is difficult to reproduce the present-day mass of the remnant.  This, in turn, affects the morphology of the Sgr stream, suppressing the spur feature along the trailing arm.  These results demonstrate that the properties of the Sgr system are sensitive to dark matter self interactions and can potentially be used to constrain SIDM models in the future, when combined with joint modeling of the host galaxy.

The paper is organized as follows. \secref{sec:methodology} discusses the methodology used to generate the simulation suites of this study, as well as algorithms used to characterize the progenitor's location and mass. \secref{sec:results} presents the key findings of the analysis, including the effects of self interactions on the remnant and stellar stream.  Finally, \secref{sec:discussion} discusses implications for applying this approach to data, and \secref{sec:conclusions} concludes the paper. Appx.~\ref{appx:scans},~\ref{appx:orbits}, and~\ref{appx:profiles} include a number of supplementary figures which give additional insight into the simulations.

\section{Methodology}
\label{sec:methodology}

\begin{table*}[ht!]
    \centering
    \begin{tabular}{l l l | c c c c c}
        \toprule
         & & & \textbf{CDM IC} & \textbf{SIDM IC} & \textbf{High Res.} & \textbf{SIDM1} & \textbf{SIDM2} \\
         \midrule
         \multirow{2}{*}{Resolution} & $m_{\DM}$ & [$10^{3}\ M_\odot$] & $875$ & $875$ & $87.5$ & $875$ & $875$  \\
         & $m_{\star}$ & [$10^{3}\ M_\odot$] & $87.5$ & $87.5$ & $8.75$ & $87.5$ & $87.5$ \\
         \midrule
         \multirow{2}{*}{Soft.~length} & $\epsilon_{\DM}$ & [pc] & 197 & 197 & 88.1 & 197 & 197 \\
         & $\epsilon_\star$ & [pc] & 98.5 & 98.5 & 44.1 & 98.5 & 98.5 \\
         \midrule
         \multirow{2}{*}{Host DM} & $M_{\DM,h}$ & $[10^{10}\ M_\odot]$ & $100$ & $100$ & $100$ & $100$ & $100$ \\
         & $a_{\DM,h}$ & $[{\rm kpc}]$ & $37$ & $37$ & $37$ & $37$ & $37$ \\
         \midrule
         \multirow{2}{*}{Host stars} & $M_{\star,h}$ & $[10^{10}\ M_\odot]$ & $7.5$ & $7.5$ & $7.5$ & $7.5$ & $7.5$ \\
         & $a_{\star,h}$ & $[{\rm kpc}]$ & $0.88$ & $0.88$ & $0.88$ & $0.88$ & $0.88$ \\
         \midrule
         \multirow{2}{*}{Sat.~DM} & $M_{\DM,s}$ & $[10^{10}\ M_\odot]$ & $1$ & $\{1, 5, 20\}$ & $1$ & $1$ & $1$ \\
         & $a_{\DM,s}$ & $[{\rm kpc}]$ & $9.0$ & $\{ 9.0, 16, 28 \}$ & $9.0$ & $9.0$ & $9.0$ \\
         \midrule
         \multirow{2}{*}{Sat.~stars} & $M_{\star,s}$ & $[10^{10}\ M_\odot]$ & $0.0555$ & $0.0555$ & $0.0555$ & $0.0555$ & $0.0555$ \\
         & $a_{\star,s}$ & $[{\rm kpc}]$ & $1.02$ & $1.02$ & $1.02$ & $1.02$ & $1.02$ \\
         \midrule
         \multirow{2}{*}{Sat.~initial vel.} & $v_i$ & [km/s] & 55--65 & 50--$\{70, 90, 160\}$ & 60 & 60 & 60 \\
         & $\theta_i$ & [deg] & 72.5--77.5 & 65--80 & 75 & 75 & 75 \\
         \midrule
         \multirow{2}{*}{Cross section} & $\sigma_0/m_\chi$ & $[{\rm cm}^2/{\rm g}]$ & 0 & 30 & $\{0,30\}$ & $\{10,20,30,40\}$ & 30 \\
         & $\eta$ & [km/s] & --- & $\infty$ & $\infty$ & $\infty$ & $\{100,200,300,400\}$ \\
         \midrule
         \multicolumn{3}{l|}{Toggle SIDM interaction type?} & --- & no & no & yes & yes \\
        \bottomrule
    \end{tabular}
    \caption{%
        An overview of the five simulation suites used in this study. The table provides the simulation particle mass for dark matter~($m_{\DM}$) and stars~($m_\star$),  the gravitational force softening length for dark matter~($\epsilon_{\DM}$) and stars~($\epsilon_\star$), the masses~($M_{(\DM/\star),(h/s)}$) and scale radii~($a_{(\DM/\star),(h/s)}$) for the Hernquist profiles used to model the dark matter and stellar density profiles of the host and satellite, the initial velocity parameters~($v_i, \theta_i$) for the satellite, and the parameters that characterize the self-interaction cross section~($\sigma_0/m_\chi, \eta$). Simulations with a velocity-independent cross section have $\sigma_0/m_\chi = \sigma/m_\chi$ and $\eta = \infty$. The last row in the table specifies whether self interactions are toggled so that they are only active for subsets of host versus satellite particles. Importantly, in all SIDM simulations, self interactions between host dark matter particles are turned off so that the host potential is comparable to CDM, which enables one to isolate the direct effects of SIDM on the Sgr system.
    } 
    \label{tab:sim_params}
\end{table*}

This section presents the methodology used in this study. \secref{sec:overview_sims} provides details of the simulations, including the codes used to compute both gravitational and SIDM interactions.  \secref{sec:satID} describes the algorithms used to assess the location and mass of the Sgr progenitor throughout its orbit. \secref{sec:scans} overviews the simulation suites and discusses the method adopted to determine the initial conditions for the progenitor.

\subsection{Simulation framework}
\label{sec:overview_sims}

All simulations in this study are run using the public version of GIZMO~\citep{gizmo}, based on the \emph{N}-body algorithms of GADGET-2~\citep{gadget2}. Self interactions are included using the SIDM implementation of \citet{sidm1} and \citet{sidm2}. The initial distributions of host and satellite simulation particles are created using GalIC~\citep{galic}, which is modified to add a random seed parameter. 

There are five separate simulation suites: 1--2) two suites (one each for CDM and SIDM) that scan over the initial phase space of the infalling satellite to reproduce the position and velocity of the Sgr dwarf today, 3) a suite of higher-resolution simulations for select parameter choices, and 4--5) two suites that vary over the SIDM model. The details of each suite are presented in \tabref{tab:sim_params}, which lists the particle resolutions, gravitational softening lengths, dark matter and stellar density profiles for the host and satellite, initial conditions for the satellite, and choices regarding the SIDM implementation.

The first row of \tabref{tab:sim_params} specifies the simulation particle mass for dark matter ($m_{\DM}$, not to be confused with the fundamental dark matter particle mass, denoted $m_\chi$) and for stars~($m_\star$). All simulations are run with $m_{\DM}=8.75 \times 10^5\,M_\odot$ and $m_\star=8.75 \times 10^4\,M_\odot$, except for the suite labeled ``High Res.'', which has $10$ times better resolution for both. The second row in the table specifies the fixed gravitational force softening lengths used in this study: $\epsilon_{\DM}$ for dark matter particles and $\epsilon_\star$ for star particles. The dark matter softening length is chosen following the \citet{powerSoftening} prescription, with $\epsilon_{\DM} \approx 4 \, r_{200} / \sqrt{N_{200, \DM}}$, where $r_{200}$ is the virial radius and $N_{200, \DM}$ is the number of dark matter particles contained within that radius. The stellar softening length is then computed as $\epsilon_\star = (m_\star / m_{\DM})^{1/3} \epsilon_{\DM} \approx 0.5 \, \epsilon_{\DM}$ to keep the mass density of each particle type approximately constant within the softening-length volume.

The next four rows of \tabref{tab:sim_params} specify the parameters for modeling the dark matter and stellar distributions of the host and satellite.  We closely follow the prescriptions of~\citet{dierickx} and point the reader to that work for extensive motivation of the parameter choices.  Here, we briefly summarize the modeling and highlight any relevant differences.

The host halo is assumed to approximately follow a Navarro-Frenk-White~\citep[NFW;][]{1997ApJ...490..493N} profile at $z=0$ with mass $\sim 10^{12}~M_\odot$ and concentration $c\sim10$.  Assuming the infall of Sgr was approximately around $z\sim 1$~\citep{dierickx}, the host would have been roughly half as massive, with a virial radius of $\sim 125$~kpc.  As our simulations do not account for the growth of the host halo, several simplifying assumptions must be made to adequately model the Sgr orbit from first infall to today: in particular, we assume that the mass accreted since $z\sim 1$ is added beyond 125~kpc and that the host's inner regions are largely unaffected.  As a result, the host can be initialized to the $z=0$ target, with the Sgr progenitor placed at $125$~kpc from the host's center at the start of its orbit. Given NFW halo parameters, GalIC produces an analogous Hernquist density profile~\citep{1990ApJ...356..359H} using a prescription detailed in \citet{galic}; the resulting total mass~($M_{\DM,h}$) and scale radius~($a_{\DM,h}$) are given in \tabref{tab:sim_params}.

While~\citet{dierickx} model the stellar profile of the host using an exponential disk and a Hernquist bulge based on~\citet{Gomez+15}, our work takes a different approach. In particular, we model the host stellar profile as a spherically symmetric Hernquist profile whose mass (enclosed within a $15$~kpc radius) is equal to the disk plus bulge system used in~\citet{dierickx}. In \tabref{tab:sim_params}, the stellar distribution parameters are indicated by the mass~($M_{\star,h}$) and scale radius~($a_{\star,h}$). The fact that the host stellar density profile has full spherical symmetry drastically simplifies the initial condition scans, which will be described in \secref{sec:scans}. This approximation is sufficient for our purposes because we are less interested in fully modeling the details of the Sgr system and more interested in comparing relative differences that arise from changing the dark matter model. 

To model the satellite, we assume a stellar Hernquist density profile with total mass $M_{\star,s} = 5.55 \times 10^{8} \, M_\odot$ and scale length $a_{\star,s} = 1.02$~kpc, chosen to be similar to the models of \citet{tango} and roughly analogous to \citet{dierickx}. Following \citet{tango}, we focus on spherical, non-rotating models for the initial Sgr progenitor; the presence of a rotating disk or substructure in the initial progenitor are thought to be important for the bifurcation of the stream arms \citep[e.g.,][]{penarrubia+2010,Oria+2022,Davies+2024} and the detailed kinematics of the present-day progenitor \citep{sgr_vasiliev}, but we consider these features out-of-scope for the present study.
For most of the simulation suites, the satellite's dark matter halo mass is $M_{\DM,s} = 10^{10} \, M_\odot$, but we have also considered $M_{\DM,s} = 5\times10^{10} \, M_\odot$ and $2\times10^{11} \, M_\odot$ for some suites.  For each mass, the NFW scale radius is set using the~\citet{dutton&maccio14} mass-concentration relation at $z=1$. \tabref{tab:sim_params} provides the corresponding scale radius assuming a Hernquist profile, $a_{\DM, s}$. We have verified that the initial density profile of the satellite is stable to self interactions. Specifically, we performed a separate test that evolved the satellite \emph{in isolation} for 3~Gyr with $\sigma/m_\chi = 30$ cm$^2$/g and did not notice any substantial changes to the density profile. This is likely due to the effect of the stellar gravitational potential on the SIDM evolution, although we have not verified this explicitly. The evolution of the dark matter density profile can be seen in \figref{fig:satellite_isolation}.

The next row of \tabref{tab:sim_params} describes the initial bulk velocity given to the satellite at infall into the Milky Way. The velocity is parameterized by its magnitude, $v_i$, and the angle $\theta_i$ made with respect to the inward radial direction (toward the center of the host). This setup is described in more detail below in \secref{sec:scans}.

The last two rows of \tabref{tab:sim_params} specify choices for the SIDM implementation of the simulation. Important inputs are the parameters of the cross section per unit dark matter mass, $m_\chi$. (Note that this is the fundamental particle mass, not the simulation particle mass.) We use the default parameterization from the public version of GIZMO, which takes the form
\begin{equation}
    \sigma(v_{\rm rel}) = \frac{\sigma_0}{1 + (v_{\rm rel} / \eta)^4} \, .
    \label{eq:SIDM_CS}
\end{equation}
Here, $v_{\rm rel}$ is the difference in velocity between the two scattering particles, $\sigma_0$ is the cross section normalization (which sets $\sigma(v_{\rm rel} \to 0)$), and $\eta$ is a velocity scale which sets the transition from constant cross section to a decreasing power law. The two free parameters that must be specified are $\sigma_0/m_\chi$ and $\eta$, and the values chosen for each suite are provided in the table. We also consider velocity-independent cross sections, which are formally equivalent to the limit $\eta \to \infty$ and thus have only one free parameter: $\sigma_0/m_\chi$ (which is equal to $\sigma/m_\chi$). Studies such as~\cite{slone+23} have shown that parameters that result in SIDM interactions within the approximate range $\sigma/m_\chi \approx 1$--$20$~cm$^2$/g at dwarf galaxy velocity scales are disfavoured by observations, while larger values of $\sigma_0/m_\chi$ with $\eta \lesssim 400$ km/s are allowed by the data.

A key goal of this study is to understand precisely \emph{how} SIDM affects the evolution of a Sgr-like system. To do this, we run a number of simulations in which we tag dark matter particles initially belonging to the satellite and particles initially belonging to the host as distinct from each other and test how the various interactions between these groups affect the system. In practice, this means only allowing scattering interactions between particles that originate from particular sources, such as satellite--satellite or satellite--host scattering events. The last row of \tabref{tab:sim_params} specifies whether or not we toggle the SIDM interaction type in each suite. Importantly, self-scattering interactions are \emph{never} allowed between two dark matter particles in the host galaxy.  This keeps the host halo potential fixed between the CDM and SIDM simulations, isolating the effects of self interactions on the Sgr dwarf and stream specifically. 
We leave a detailed study of the SIDM impacts on the host halo---and the additional consequences that has on the Sgr system---to future work.  

\subsection{Satellite identification}
\label{sec:satID}

An important element of this analysis is to track the evolution of the satellite progenitor along its orbit.  Due to the extreme tidal forces that act on the satellite, we find that standard algorithms \citep[such as a simple shrinking sphere or the AMIGA Halo Finder,][]{AHF} struggle to identify the progenitor to the present day.  As such, we use the DBSCAN algorithm as implemented in \texttt{scikit-learn}~\citep{dbscan,sklearn}. DBSCAN identifies clusters in a set of points according to their spatial density. It is robust to noise and can classify points that do not belong to any cluster as background. The algorithm depends on a parameter (denoted \verb+eps+), which sets the maximum distance between points that can be considered in the same neighborhood. We assume that among the satellite stellar particles, the progenitor is the point of largest density, so we iteratively increase \verb+eps+ until only one cluster is identified and then take the stars identified in this way to be the progenitor. The position and velocity of the progenitor is computed as the mean position and velocity of all identified stars.

This algorithm is used \textit{only} to determine the location and velocity of the progenitor. The mass of the satellite is estimated by summing the masses of all particles within a $5$~kpc sphere centered at the progenitor's inferred position, considering only particles that are initialized as satellite particles. This sum includes particles that are tidally stripped from the satellite at some earlier time, but happen to be within $5$~kpc of the progenitor's location. To remedy this, we calculate the average mass density of satellite particles within a shell-shaped volume that is centered on the host and whose inner and outer radii are $5$ kpc smaller and larger than the satellite's location (the spherical volume surrounding the satellite used for the initial sum is not included when calculating the average density). This average density is then used to estimate the background mass of unbound particles in the $5$ kpc sphere surrounding the satellite's location and that background is removed from the mass estimate of the satellite.

\subsection{Overview of simulation suites}
\label{sec:scans}

\noindent \textbf{CDM and SIDM IC Suites.} The goal of these two suites is to find initial conditions for the progenitor that evolve to match the position and velocity of the Sgr dwarf today. The final (target) Galactocentric phase-space coordinates of the Sgr dwarf taken from~\citet{tango} are $\vec{x}_{\text{tgt}} = (17.9, 2.6, -6.6)~\text{kpc}$ and $\vec{v}_{\text{tgt}} = (239.5, -29.6, 213.5)~\text{km/s}$. This target is taken to be the satellite position after three pericentric passages, as motivated by \citet{fardal+19}. The CDM~IC suite attempts to find the phase-space coordinates of the satellite's first infall that reproduce $\{\vec{x}_{\text{tgt}},\,\vec{v}_{\text{tgt}}\}$, assuming a CDM model with $M_{\DM,s} = 10^{10} \ M_\odot$. The SIDM~IC suite does the same but for an SIDM model with a velocity-independent cross section of $\sigma/m_\chi = 30 \ \text{cm}^2/\text{g}$ and three variations of the satellite's infall mass $M_{\DM,s}=\{10^{10}, 5\times 10^{10}, 2 \times 10^{11}\} \, M_\odot$.

For all simulations, the progenitor is initiated at $125$~kpc from the host's center and a velocity vector is chosen by specifying its magnitude, $v_i$, and the angle between the velocity vector and the radial from the host center, $\theta_i$ (defined such that $\theta_i = 0^\circ$ corresponds to a radially inward velocity). The simulation is then run and the orbit of the progenitor, $\{\vec{x}_j,\vec{v}_j\}$, is tracked throughout the simulation at timesteps, denoted by a subscript $j$, with $10$~Myr spacings. The goal is to find the set of $\{v_i,\theta_i\}$ that result in phase-space values close to $\{\vec{x}_{\text{tgt}},\,\vec{v}_{\text{tgt}}\}$ after three pericenters.

Because our model of the Milky Way is spherically symmetric, it is always possible to align the orbital plane of the simulation with the plane defined by the vectors $\{\vec{x}_{\text{tgt}},\,\vec{v}_{\text{tgt}}\}$. Then, the only ambiguity in comparing $\{\vec{x}_j,\vec{v}_j\}$ to $\{\vec{x}_{\text{tgt}},\,\vec{v}_{\text{tgt}}\}$ is an arbitrary rotation around the axis perpendicular to the orbital plane. In order to set this angle, at every snapshot between the third and fourth pericenter, we align the position of the satellite $\vec{x}_j$ to be on the same radial direction as the vector $\vec{x}_{\text{tgt}}$. This is motivated by the small uncertainties in the on-sky position of Sgr (however, we note that our mock observations are effectively from the Galactic Center). Then, for each snapshot, we define the chi-squared value to be
\begin{equation} \label{eq:chi2}
\begin{aligned}
    \chi^2_j 
    & \equiv \frac{\| \vec{x}_j - \vec{x}_{\text{tgt}}\|^2}{\sigma_x^2}
    + \frac{\| \vec{v}_j - \vec{v}_{\text{tgt}}\|^2}{\sigma_v^2} \\
    & = \frac{(r_j - r_{\text{tgt}})^2}{\sigma_x^2} + \frac{v_j^2 + v_{\text{tgt}}^2 - 2v_j v_{\text{tgt}} \cos(\theta_j - \theta_{\text{tgt}})}{\sigma_v^2} \, ,
\end{aligned}
\end{equation}
where the weights are taken to be $\sigma_x = 0.1 \ \text{kpc}$ and $\sigma_v = 1 \ \text{km/s}$, motivated by observational uncertainties and the choices made in~\citet{tango}. If the progenitor is fully disrupted before snapshot $j$, then $\chi^2_j$ is set to $\infty$. Then, for each simulation, we define $\chi^2 \equiv \min_j \chi^2_j$ to be the minimum $\chi^2_j$ that occurs between the third and fourth pericenter.

The second row in~\eqref{eq:chi2} is the explicit calculation of $\chi_j^2$ after the rotation described above is performed. In this case, the numerator of the first term reduces to the square of the difference in radii between the satellite and the radius of the target location, where $r_{j/{\rm tgt}} \equiv |\vec{x}_{j/{\rm tgt}}|$. For the second term, we defined 
$v_{j/{\rm tgt}} \equiv |\vec{v}_{j/{\rm tgt}}|$ and $\cos{(\pi-\theta_{j/{\rm tgt}})} \equiv (\vec{x}_{j/{\rm tgt}} \cdot \vec{v}_{j/{\rm tgt}}) / (r_{j/{\rm tgt}} v_{j/{\rm tgt}})$. Note that $v_{j=0}$ and $\theta_{j=0}$ defined in this way match the definitions of $v_i$ and $\theta_i$.

Scanning over $v_i$ and $\theta_i$ is done by constructing a grid with spacings $10 \ \text{km/s}$ and $5^\circ$ and performing simulations at these grid points. After each simulation is complete, we find the grid point with minimal $\chi^2$. If it is surrounded on all sides by simulations with larger values of $\chi^2$, we conclude the scan; otherwise, we run additional simulations to expand the grid towards the minima of $\chi^2$. For the CDM scan, we enhanced the grid-space resolutions to $2.5 \ \text{km/s}$ and $1.25^\circ$. The final ranges of the scanned parameters are listed in \tabref{tab:sim_params}.

For the CDM~IC suite, the scan converges on $\{ v_i, \theta_i \} = \{ 60 \ \text{km/s}, 75^\circ \}$. The closest point of comparison for these values is the \citet{dierickx} semi-analytic scan, which found a best-fit at $\{ v_i, \theta_i \} = \{ 80 \ \text{km/s}, 80^\circ \}$ using heavier Sgr and Milky Way (total halo masses 30\% and 25\% larger) and evolving an additional two pericentric passages. 

For the SIDM~IC suite with $M_{\DM,s} = 10^{10} \ M_\odot$, the scan also converges on $\{ v_i, \theta_i \} = \{ 60 \ \text{km/s}, 75^\circ \}$. For $M_{\DM,s} = 5 \times 10^{10} \ M_\odot$, the scan converges on $\{ v_i, \theta_i \} = \{ 80 \ \text{km/s}, 75^\circ \}$. For $M_{\DM,s} = 2 \times 10^{11} \ M_\odot$, the scan nearly converges on $\{ v_i, \theta_i \} = \{ 100 \ \text{km/s}, 65^\circ \}$. We stop this scan early because the fits are poor in this case. 

See \appref{appx:scans} for visualizations of the $\chi^2$ values as a function of grid position.

\vspace{1em}
\noindent\textbf{High Resolution Suite.} This suite consists of two high-resolution simulations that are run using the best-fit initial conditions of the CDM~IC and SIDM~IC suites. The resolution of these simulations is $10$ times that used in the initial conditions scans, which corresponds to simulation particle masses of $m_{\DM} = 8.75 \times 10^4 \, M_\odot$ and $m_\star = 8.75 \times 10^3 \, M_\odot$. The SIDM simulation includes both satellite--satellite and satellite--host interactions. The purpose of these high-resolution runs is to better recover details of the stream morphology for a few model benchmarks.

\vspace{1em}
\noindent\textbf{SIDM1 and SIDM2 Suites.} The goal of these two simulation suites is to test how various aspects of the SIDM model affect a Sgr-like system. In particular, we test variations to the SIDM cross section (including its velocity dependence) and also perform a novel test of how satellite--host and satellite--satellite interactions affect the evolution of the system.

The SIDM1 suite includes a set of velocity-independent simulations, with the value of the cross section corresponding to $\sigma/m_\chi = 10,\,20,\,30$ and $40$ cm$^2$/g. The SIDM2 suite includes SIDM simulations with a constant value of $\sigma_0/m_\chi = 30$ cm$^2$/g and different velocity dependencies corresponding to $\eta = 100,\,200,\,300$ and $400$~km/s. For both suites, the satellite halo mass is taken to be $10^{10} \, M_\odot$ and the initial conditions are set to the best-fit values found in the corresponding SIDM~IC suite described above. Using the same initial velocity values for these cross-section variations is an adequate approximation; for all parameters, we match the present-day position to within $2.1 \ \text{kpc}$ and the velocity to within $40 \ \text{km/s}$ (with most at less than $25 \ \text{km/s}$). For more detail, see \appref{appx:orbits} and in particular \figref{fig:vary_orbits}.

\begin{figure*}
    \centering
    \includegraphics[width=\linewidth]{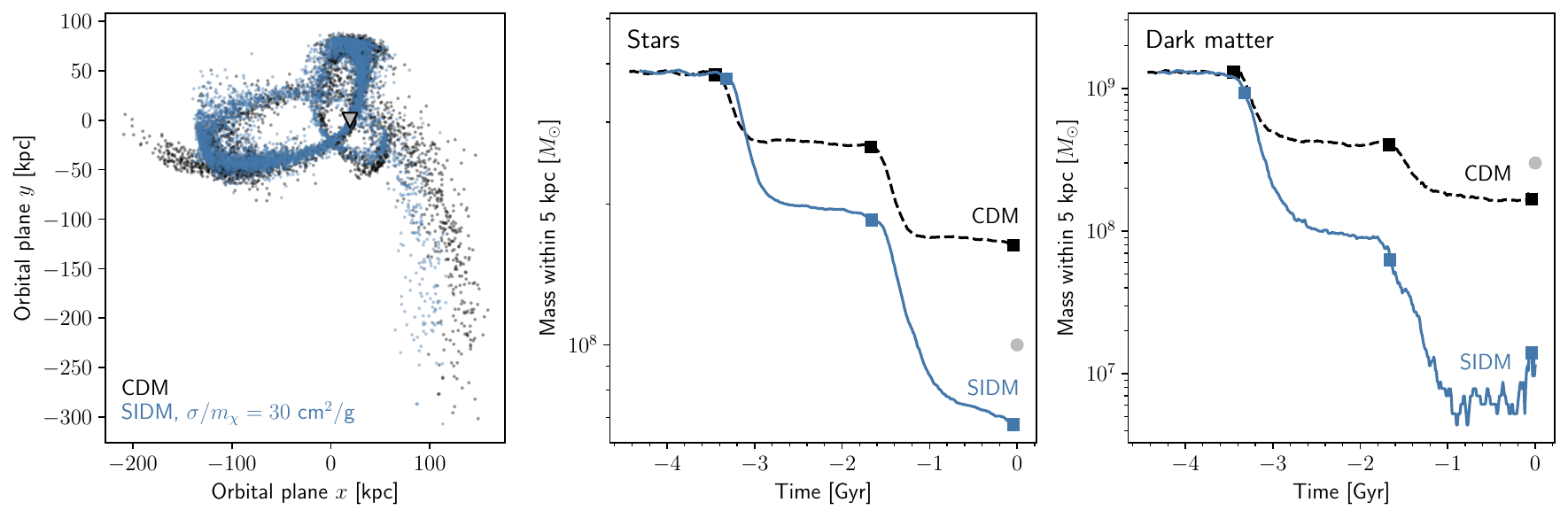}
    \caption{%
        Results from the best-fit simulations of the CDM~(black) and SIDM~(blue) IC suites, for the $M_{\DM, s} = 10^{10}~M_\odot$ progenitor.  The SIDM model assumes a velocity-independent cross section of $\sigma/m_\chi = 30 \text{ cm$^2$/g}$. The left panel shows the final positions of the satellite star particles in the orbital plane, rotated to place the progenitor on the $x$-axis. The middle and right panels show the mass from stars and dark matter enclosed within 5~kpc of the progenitor position as a function of time, with pericenter times marked by squares. The grey triangle (with black outline) in the left panel marks the target position of the progenitor. The grey circle in the middle~(right) panel is an estimate of the enclosed stellar~(dark matter) mass of Sgr from \citet{sgr_vasiliev}, which is included to give a reference for the anticipated orders-of-magnitude of these masses. This SIDM model exhibits significantly enhanced mass loss relative to CDM in stars and especially in dark matter. 
    }
    \label{fig:scan_best_2x}
\end{figure*}

\section{Results}
\label{sec:results}

This section reviews the primary results regarding the Sgr progenitor mass~(\secref{sec:massloss}) and stream morphology~(\secref{sec:stream}).  As a concrete example, we compare CDM with the case of a velocity-independent SIDM cross section of $\sigma/m_\chi = 30$~cm$^2$/g in these two subsections. \secref{sec:SIDM_variations} then  describes how the key results change when varying the SIDM parameters.

\subsection{Progenitor mass}
\label{sec:massloss}

The left panel of \figref{fig:scan_best_2x} shows the stellar debris from the best-fit simulations of the CDM and SIDM~IC suites, for the $M_{\DM, s} = 10^{10}~M_\odot$ progenitor.  As viewed in the orbital plane, a clear stellar stream forms in both models. The present-day location of the Sgr dwarf ($\vec{x}_{\mathrm{tgt}}$) in this frame is indicated by the grey triangle; this point occurs just after the satellite's third pericentric passage. 

A striking difference between the CDM and SIDM simulations is the evolution of the satellite's mass, shown in the central and right panels of \figref{fig:scan_best_2x} for stars and dark matter located within 5~kpc of the satellite's center. As expected, the mass evolution for both stars and dark matter exhibits long plateaus punctuated by sharp drops at pericenter where disruption is strongest. For the CDM simulation, the satellite mass at present is approximately $\sim 2 \times 10^8 \ M_\odot$ for both stars and dark matter. To within a factor of $\sim 2$, the CDM simulation reproduces the estimated enclosed stellar mass of $\sim 10^{8}\ M_\odot$ and dark matter mass of $\sim 3 \times 10^{8}\ M_\odot$ of Sgr \citep[taken from dynamical modeling of the progenitor based on observations,][]{sgr_vasiliev}, which are indicated by the grey points in the figure. 

In the case of SIDM, the satellite exhibits significantly larger mass drops at each pericenter and more prolonged periods of mass loss. For example, during its first pericentric passage, the CDM satellite loses approximately $60\%$ of its dark mass while the SIDM satellite loses approximately $90\%$. At present day, the SIDM satellite has an enclosed stellar mass within 5~kpc of  $\sim 7 \times 10^7 \ M_\odot$; the corresponding dark matter mass is $\sim 1 \times 10^7 \ M_\odot$. Note that the dark matter mass loss is so significant that by the end of the simulation there are only $\mathcal{O}(10)$ dark matter particles near the progenitor center. In this case, the stellar mass is a factor of $\sim 1.4$ smaller than the observations, but the dark matter mass is more than an order-of-magnitude smaller.

\begin{figure*}
    \centering
    \includegraphics[width=0.8\linewidth]{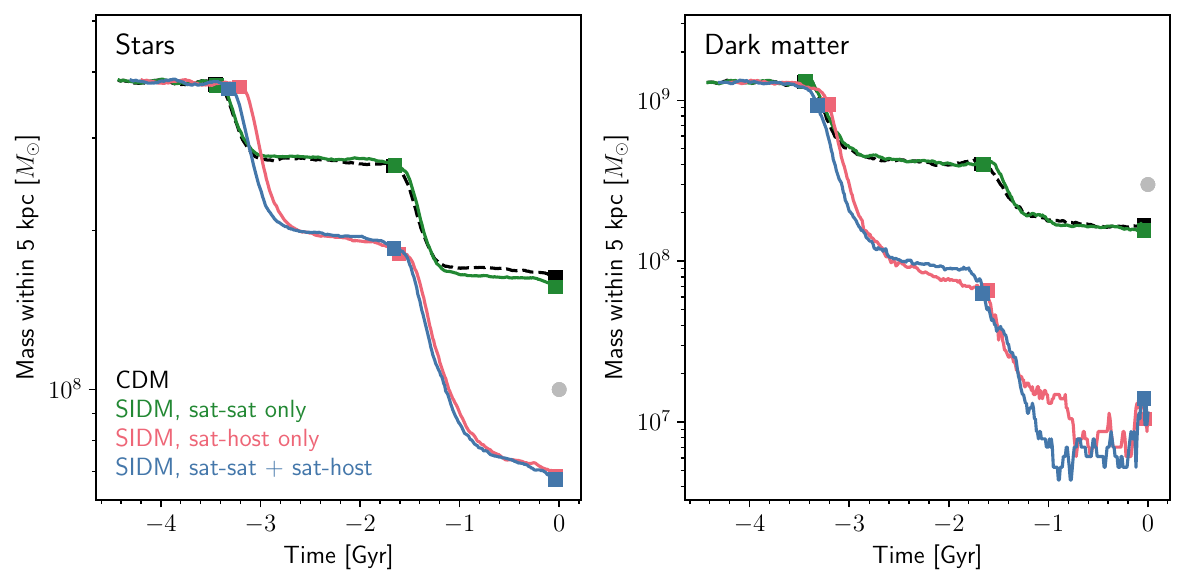}
    \caption{%
        The stellar (left) and dark matter (right) mass enclosed within 5 kpc of the progenitor for the best-fit simulations of the CDM and SIDM~IC suites with $M_{\DM, s} = 10^{10}~M_\odot$. Squares mark the times of pericenters. The `CDM' simulation~(black) has no dark matter interactions; `sat--sat only'~(green) has interactions between satellite dark matter particles; `sat--host only'~(pink) has interactions between satellite and host dark matter particles; and `sat--sat + sat--host'~(blue) has both satellite--satellite and satellite--host interactions. All SIDM interactions here assume a velocity-independent cross section of $\sigma/m_\chi = 30$ cm$^2$/g. The grey circle in each panel is an estimate of the enclosed stellar/dark matter mass of Sgr from \citet{sgr_vasiliev}. There is little difference between CDM and sat--sat only, and little difference between sat--host only and sat--sat + sat--host, indicating that ram-pressure evaporation is responsible for the enhanced mass loss observed for this SIDM model.
    }
    \label{fig:isolations}
\end{figure*}

As previously discussed, there are two primary mechanisms by which dark matter self interactions can affect mass loss: the first is through the satellite's central density, which affects tidal disruption, and the second is through ram-pressure evaporation. To better understand the relative contribution of each mechanism, we modify the SIDM implementation to control which dark matter particles are allowed to interact. In one case, only satellite--satellite interactions are allowed and in the other, only satellite--host interactions. The results for the stellar and dark matter mass loss in these simulations are shown in \figref{fig:isolations}. The black~(blue) curve is  a reproduction of the CDM~(SIDM) curve from \figref{fig:scan_best_2x}.  Additionally, the result of only satellite--satellite interactions is shown in green and only satellite--halo interactions is shown in pink. The former has nearly identical mass evolution to the CDM simulation at all times, demonstrating that the self interactions internal to the satellite do not have a significant effect on its tidal history. When the satellite--host interactions are turned on, however, there is significant variation from the CDM case.  These results demonstrate that the mass-loss history for SIDM can be explained by ram-pressure evaporation.

The rate of ram-pressure evaporation can be approximately described by~\citep{sidm_rampressure,slone+23}
\begin{equation}
\Gamma_{\rm rp} 
\approx 4\,{\rm Gyr}^{-1} 
\times 
\left(
    \frac{\rho_{\rm host}(\vec{x})}{2 \times 10^6\,\frac{M_\odot}{{\rm kpc}^3}}
\right) 
\times 
\left(
    \frac{
        \langle\chi_e \, \tilde{\sigma}/m_\chi \, v_{\rm rel}\rangle_{\rm orb}
    }{
        30 \frac{{\rm cm}^2}{{\rm g}} \, \cdot \, 
        300 \frac{{\rm km}}{{\rm s}}
    }
\right) 
,
\label{eq:rp}
\end{equation}
where $\rho_{\rm host}(\vec{x})$ is the mass density of the host halo at the location of the orbiting satellite, $\chi_e$ is a unitless prefactor that is $\mathcal{O}(1)$ for $v_{{\rm esc},s} \ll \text{min}[\eta,\, v_{\rm rel}]$ with $v_{{\rm esc},s}$ the satellite's escape velocity, $\langle...\rangle_{\rm orb}$ refers to averaging over velocity differences between satellite and host particles at the location of the satellite, and $\tilde{\sigma}$ is the angle-integrated SIDM cross section. The instantaneous mass loss due to ram pressure is then approximately given by
\begin{equation}
\frac{\dot{M}_{\DM,s}}{M_{\DM,s}} \approx - \Gamma_{\rm rp}.
\label{eq:rp_2}
\end{equation}
For $\tilde{\sigma} \approx 30$~cm$^2$/g and for typical host densities and orbital velocities experienced by the Sgr dwarf, the numerical rate evaluated in~\eqref{eq:rp} indicates substantial mass loss via ram-pressure evaporation over sub-Gyr timescales, in agreement with our simulation results.

For the simulations presented thus far, the halo mass of Sgr is initialized to $M_{\DM, s} = 10^{10}$~M$_\odot$.  A natural question is whether increasing the initial mass may counteract the enhanced mass loss in SIDM and result in a present-day mass that better matches observations. To test this, we have performed two additional SIDM initial conditions scans with satellite halo masses of $5 \times 10^{10} \ M_\odot$ and $2 \times 10^{11} \ M_\odot$, roughly spanning the upper range of potential Sgr progenitor masses quoted in the literature \citep[e.g.,][]{gibbons+2017, jiang&binney}.\footnote{%
    For these simulations, the stellar mass, $M_{\star, s}$, is initialized at the same value as for the $M_{{\DM},s}=10^{10}\,M_\odot$ simulation since the main purpose of these simulations is to test conditions that might increase the final dark matter mass of the remnant.
} The latter case is admittedly quite extreme as the progenitor's mass is 20\% of the host's mass at infall. 

\begin{figure*}
    \centering
    \includegraphics[width=\linewidth]{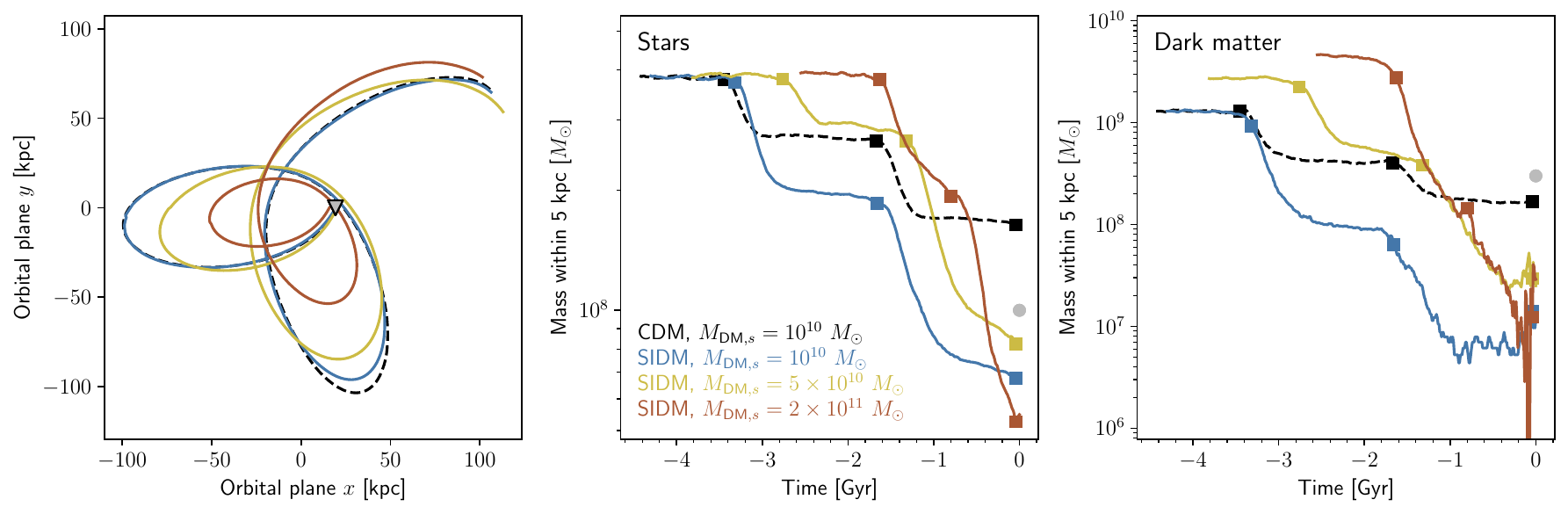}
    \caption{%
        Results from the best-fit simulations of the CDM and SIDM~IC suites, varying the initial dark matter mass of the Sgr progenitor.  All SIDM interactions here assume a velocity-independent cross section of $\sigma/m_\chi = 30$ cm$^2$/g. The left panel shows the trajectories of the progenitors in the orbital plane, rotated to place the present-day position on the $x$-axis. The grey triangle (with black outline) marks the target position of the progenitor. The middle (right) panel shows the stellar (dark matter) mass enclosed within 5~kpc of the progenitor, with squares marking the times of pericenters and a grey circle marking the observed estimates of these quantities from \citet{sgr_vasiliev}. The black and blue curves correspond to the original CDM and SIDM simulations, provided in \figref{fig:scan_best_2x}, with  initial halo mass $M_{\DM,s} = 10^{10} \ M_\odot$.  The yellow and brown curves show the results for SIDM when the initial halo mass is increased to $M_{\DM,s} = 5 \times 10^{10}$ and $2 \times 10^{11}  \ M_\odot$, respectively, keeping the initial stellar mass of the satellite constant.  The SIDM simulations do not reproduce the final dark matter mass of the satellite seen in CDM simulations, even when the progenitor is initialized with more mass.}
    \label{fig:heavy}
\end{figure*}

\figref{fig:heavy} shows the orbital trajectories and enclosed mass histories for these more massive progenitors. The orbital trajectories in the left panel demonstrate that increasing the mass of the progenitor leads to drastically different orbital characteristics after completing a scan over initial conditions. The heavier satellites prefer less tangential initial orbits (for halo masses $\{ 1, 5, 20 \} \times 10^{10} \ M_\odot$, $\theta_i = \{ 75, 75, 65 \}$~deg) and larger initial velocities ($v_i = \{ 60, 80, 100 \}$~km/s). This in turn leads to smaller apocenters (around $\{ 100, 80, 55 \}$~kpc) and shorter orbital periods (around $\{ 1.6, 1.3, 0.8 \}$~Gyr).

Despite these differences, the more-massive progenitors still exhibit significant ram-pressure evaporation, as shown in the central and right panels of \figref{fig:heavy}. Indeed, negligibly small amounts of dark matter remain by the third pericentric passage, regardless of the initial halo mass of the progenitor.\footnote{ Especially for the most massive cases, the satellite rapidly decays to the center of the host, dumping a considerable amount of dark matter there that is then challenging to distinguish from the remnant at last pericenter. This results in some additional noise in the enclosed mass estimate, but does not affect the overall conclusion that basically no dark matter remains bound at late times.}  This demonstrates just how difficult it is to recover the correct present-day mass of Sgr in the presence of strong self interactions (specifically, velocity-independent $\sigma/m_\chi = 30 \text{ cm}^2/\text{g}$).

\subsection{Stream morphology}
\label{sec:stream}

Next, we consider the properties of the stellar stream itself.  The results presented here use simulations from the High Resolution suite. The order-of-magnitude improvement in particle-mass resolution for the dark matter and stars better captures the details of the stream morphology. As mentioned in \secref{sec:methodology}, both the CDM and SIDM high-resolution simulations use the best-fit initial conditions obtained from their lower-resolution counterparts in the CDM and SIDM~IC suites. 

\begin{figure*}[h]
    \centering
    \includegraphics[width=0.9\linewidth]{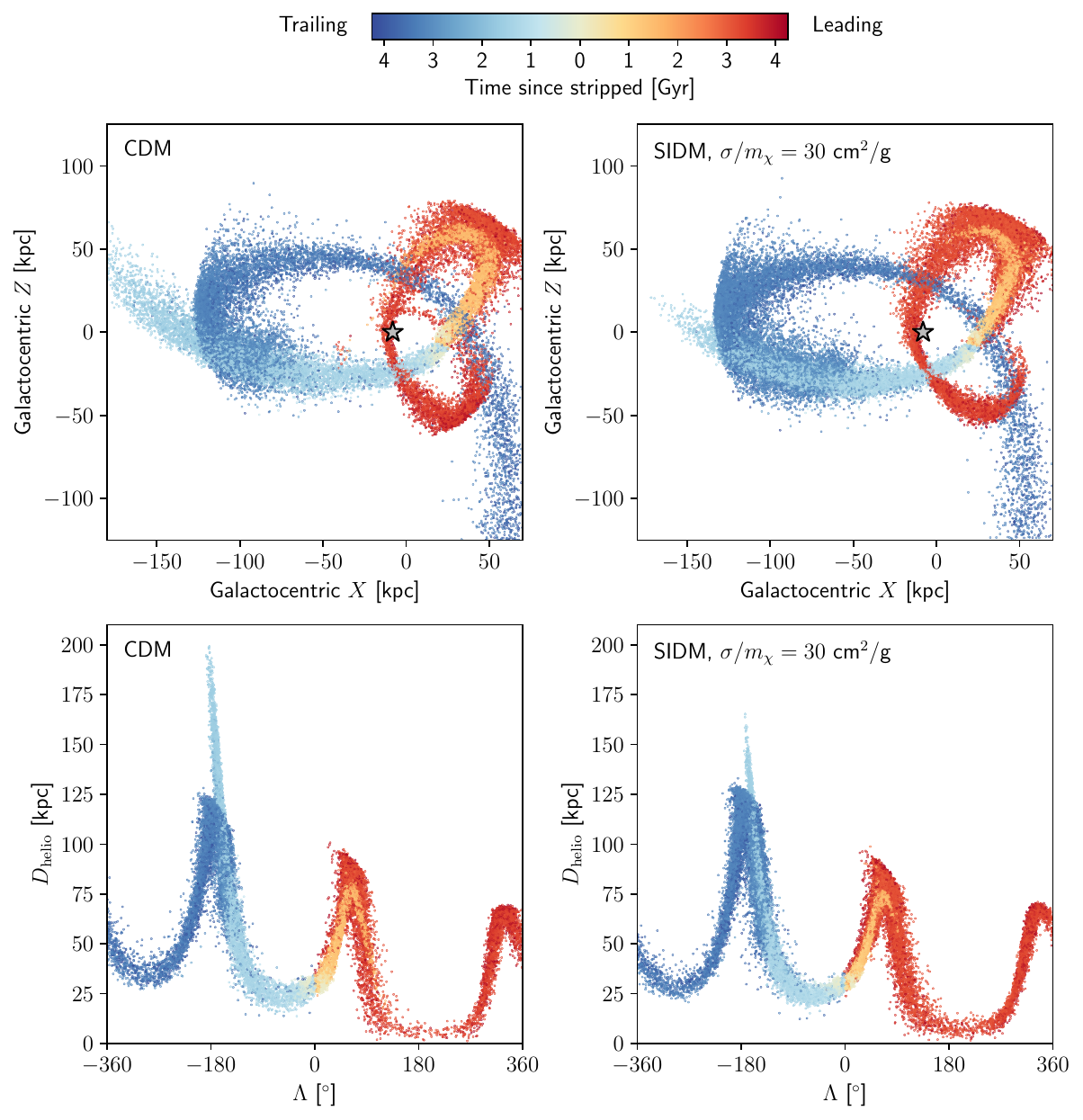}
    \caption{%
        The stellar stream from the High Resolution suite for CDM~(left) and SIDM~(right). The SIDM interactions assume a velocity-independent cross section of $\sigma/m_\chi = 30$ cm$^2$/g. The top row shows the stream in the $XZ$-plane of the Galactocentric coordinate system, with the Milky Way disk in the $XY$-plane and the Sun located at $(-8.122, 0, 0.0208)$~kpc (marked by the grey star) \citep{astropy:2022}. The bottom row shows the heliocentric distance of the stream particles as a function of heliocentric stream angle, $\Lambda$. The distributions are markedly similar, with the largest difference being the population of light blue, recently stripped stars at $\Lambda \sim -170^\circ$ attaining significantly larger distances in CDM than in SIDM.
    }
    \label{fig:stream_compare}
\end{figure*}

\begin{figure*}[t!]
    \centering
    \includegraphics[width=0.8\linewidth]{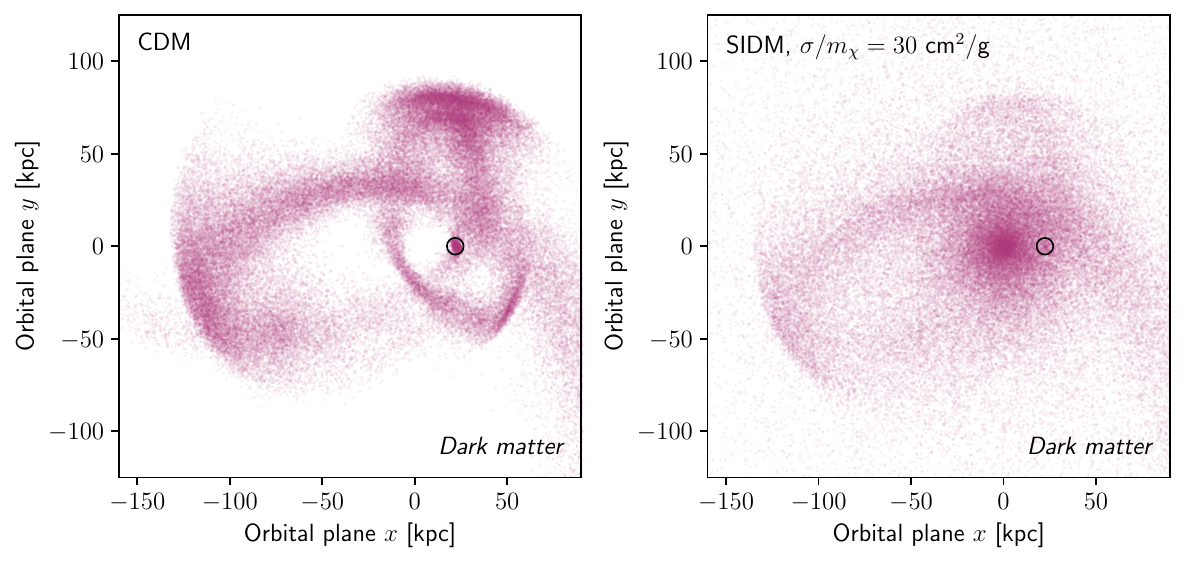}
    \caption{%
        The final positions of dark matter particles in the High Resolution simulations for CDM~(left) and SIDM~(right). The SIDM interactions assume a velocity-independent cross section of $\sigma/m_\chi = 30 \ \text{cm}^2/\text{g}$. The positions are shown in the orbital plane, rotated to place the progenitor (marked by a circle) on the $x$-axis. In CDM, the dark matter largely traces the stars, whereas in SIDM, the stream and progenitor are highly disrupted and most of the stripped dark matter is concentrated in the center of the host halo.%
    }
    \label{fig:dm_stream}
\end{figure*}

\figref{fig:stream_compare} shows the stellar stream that forms for the CDM~(left) and SIDM~(right) simulations. The top row shows the stream in the $XZ$-plane of the Galactocentric coordinate system, with the Milky Way disk in the $XY$-plane, $Z$ aligned with the angular momentum of the disk, and the Sun located at $(-8.122, 0, 0.0208)$~kpc~\citep{astropy:2022}.
The bottom row provides the heliocentric distance of simulated Sgr stream stars, $D_{\rm helio}$, as a function of the stream angle $\Lambda$, which is computed using the rotation matrix provided in \citet{belokurov+14}. During the simulation, we track $\Lambda$ for all particles, adding or subtracting $2 \pi$ whenever a particle crosses $\Lambda = \pm \pi$, so that the stream is entirely ``unwrapped''. For each particle shown in the figure, the last time at which that particle was within 5~kpc of the progenitor is taken as a proxy for the time since it was unbound from the satellite. Those times are represented by the coloring of each point, with a different color scale chosen for particles belonging to the leading versus trailing arms of the stream. 

The left panels of \figref{fig:stream_compare} show the stellar stream that forms in the CDM simulation. The stream qualitatively resembles the observed Sgr stream in many ways. With the progenitor located at $\Lambda = 0^\circ$, the leading and trailing arms extend to positive and negative $\Lambda$, indicated in red and blue, respectively. The apocenters of the leading and trailing arms are enhanced relative to observations (e.g.,~leading $47.5 \pm 1.4 \ \text{kpc}$, trailing $92.7 \pm 1.3 \ \text{kpc}$ \citep{hernitschek+17}), but the trailing arm apocenter is characteristically larger. The simulations also reproduce the spur on the apocenter of the trailing arm; this is the dramatic spike of light blue, recently stripped stars at $\Lambda \sim -170^\circ$. This spur has been observed in other CDM simulations of the Sgr stream, including those by~\citet{2015MNRAS.452..301F,dierickx,gibbons+2017, fardal+19}, as well as in data~\citep{spur_sesar+17,2019MNRAS.490.5757S, Bayer+2025}.

The right panels of \figref{fig:stream_compare} show the stellar stream that forms in the SIDM simulation.  Overall, the stream resembles that formed in CDM, including in the location of its pericenters and apocenters. The most significant difference between the streams is in the spur, which attains significantly larger heliocentric distances for the CDM simulation ($\sim 200$~kpc) compared to the SIDM simulation ($\sim 170$~kpc). The heliocentric distance of the spur is an indicator of the energy of stars in the progenitor at the time those stars were stripped, which is the previous pericenter. The larger the energy of those stripped stars, the larger the value of $D_{\rm helio}$ they can attain. Stars with energies larger than the orbital energy of the progenitor drift in the direction of the trailing arm of the stream, and vice versa. Thus, the large $D_{\rm helio}$ spur in the trailing arm consists of the highest energy stars stripped at the previous pericenter overtaking stars stripped two pericenters ago. There is a qualitatively similar feature in the leading arm due to the lowest energy stars stripped at the previous pericenter overtaking stars stripped two pericenters ago. These stars have a smaller apocenter and do not stand out as clearly from the previously stripped material (see \figref{fig:stream_compare}).

Since the velocity dispersion of stars is a proxy for the depth of the gravitational potential in which they reside, the height of the spur is a proxy for the amount of dark matter enclosed within the stellar region of the progenitor during the last pericentric passage, which in turn depends on the rate of mass loss during previous pericentric passages. Therefore, simulations with prominent and high $D_{\rm helio}$ spurs should correspond to progenitors with little mass loss from their central regions, while progenitors with significant mass loss should decrease the stellar population's velocity dispersion and decrease the visibility of a spur. Dark matter with self interactions large enough to efficiently remove mass via ram-pressure evaporation would be particularly effective at suppressing the spur since such interactions are able to remove mass from the most central regions of a satellite. This is in contrast to mass removal via tidal stripping, which initially removes mass from the outskirts of the satellite.

We also consider the morphology of the dark matter stream that forms in these simulations. The final distribution of dark matter particle positions is shown in \figref{fig:dm_stream} for CDM (left) and SIDM (right). In CDM, the dark matter closely follows the distribution of stars, with particles either bound to the progenitor or dispersed along the stellar stream. Strikingly, SIDM does not show this; instead, much of the dark matter is concentrated in the central region of the host and is more phase mixed. This could be a result of ram-pressure evaporation, which predominantly removes mass during pericentric passages and also transfers substantial momenta to the evaporated particles.

\begin{figure*}[t]
    \centering
    \includegraphics[width=0.8\linewidth]{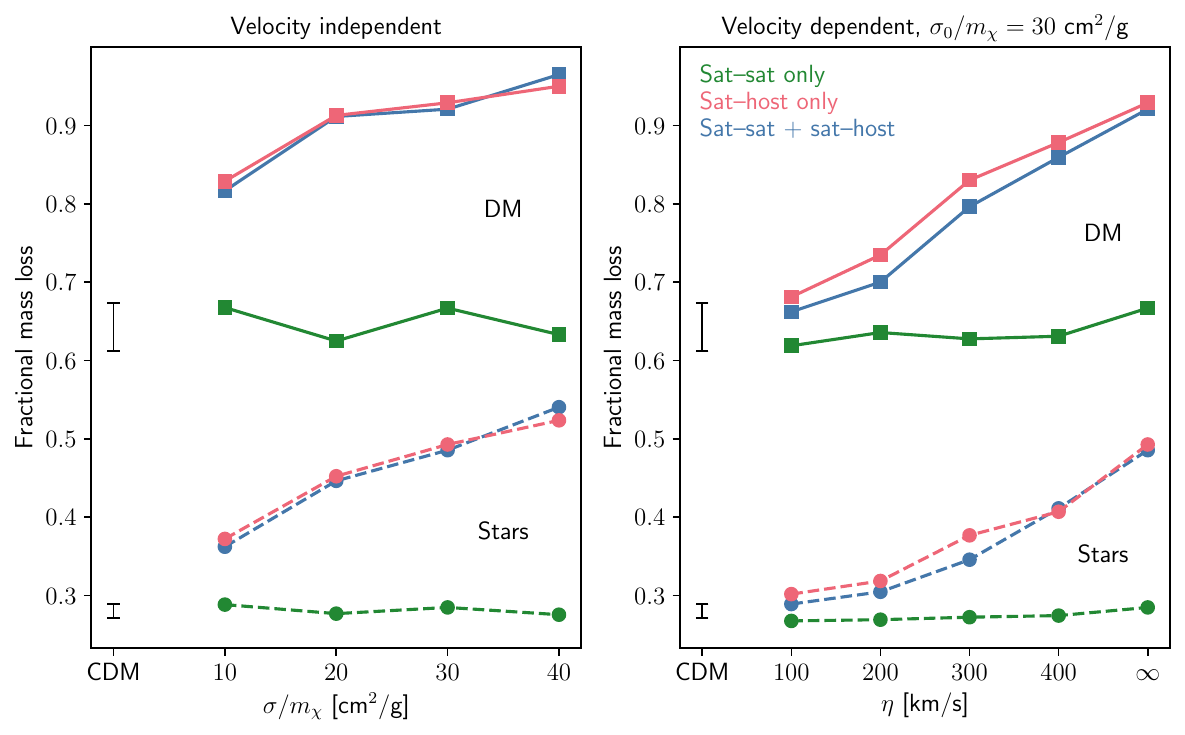}
    \caption{%
        Fraction of stellar (circles and dashed lines) and dark matter (squares and solid lines) mass lost between initialization and the first apocenter. The left panel shows the mass loss as a function of the velocity-independent cross section $\sigma/m_\chi$ and the right panel as a function of the velocity scale $\eta$ for a velocity-dependent cross section (using $\sigma_0/m_\chi = 30$ cm$^2$/g). Colors indicate the allowed types of self interactions, with green being satellite--satellite only, pink being satellite--host only, and blue being both. In black is shown the range of mass losses observed from 11 CDM simulations varying the random seed of the initial conditions generator. There is little difference between our full SIDM simulations and the satellite--host only simulations and little difference between satellite--satellite only simulations and the CDM simulations, which indicates that the enhanced SIDM mass loss is driven predominantly by ram-pressure evaporation at all model points.
    }
    \label{fig:vary_both_isolations}
\end{figure*}

\subsection{Varying SIDM parameters}
\label{sec:SIDM_variations}

\begin{figure*}[t]
    \centering
    \includegraphics[width=0.9\linewidth]{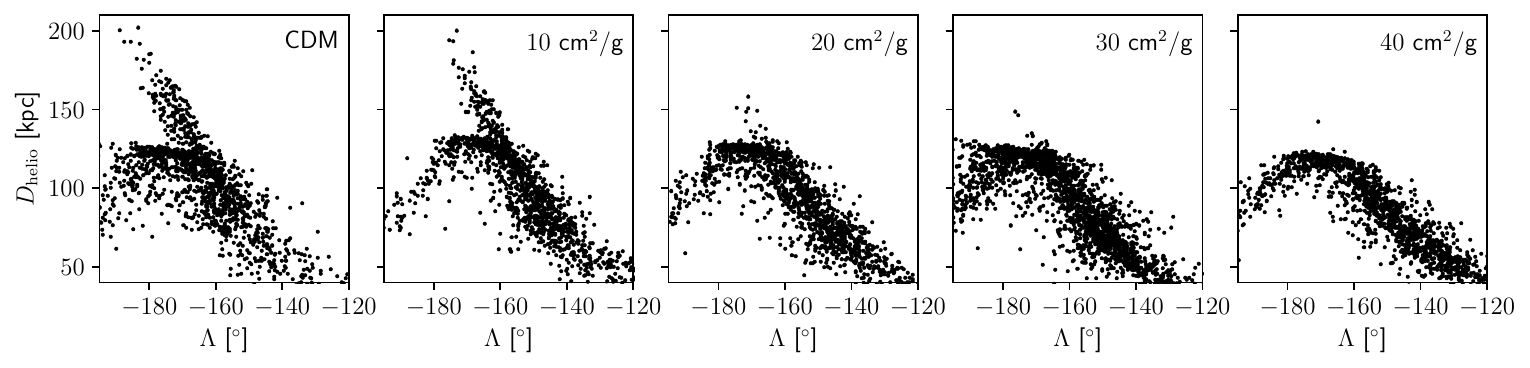}
    \includegraphics[width=0.9\linewidth]{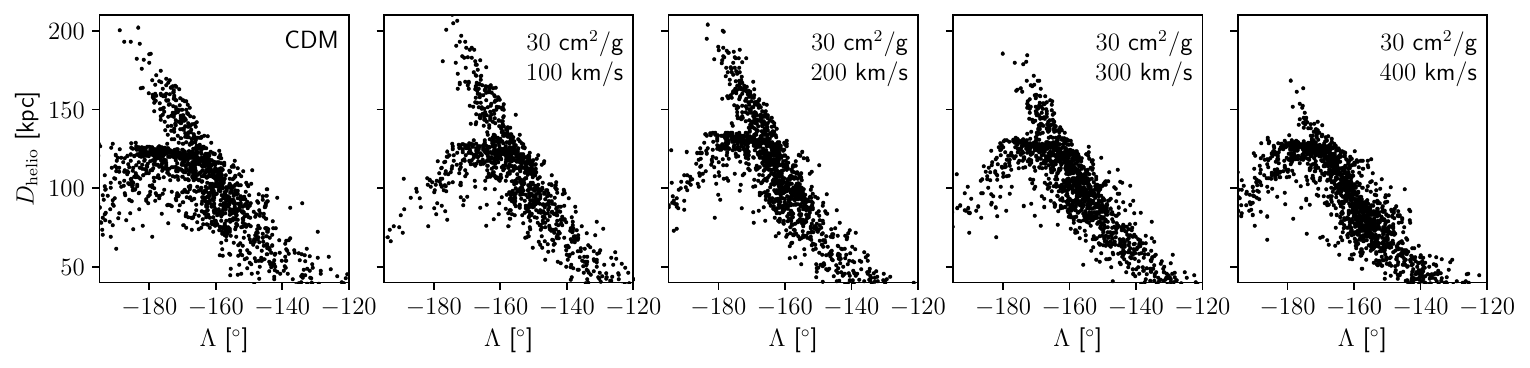}
    \caption{%
        Heliocentric distance $D_{\text{helio}}$ versus stream angle $\Lambda$ for the spur region of the stellar stream for standard-resolution simulations varying the SIDM model parameters. The top row shows the CDM simulation alongside velocity-independent simulations with cross sections $\sigma/m_\chi = \{ 10, 20, 30, 40 \} \text{ cm}^2/\text{g}$. The bottom row shows the CDM simulation again alongside velocity-dependent simulations with cross section $\sigma_0/m_\chi = 30 \text{ cm}^2/\text{g}$ and velocity scale $\eta = \{ 100, 200, 300, 400 \} \text{ km}/\text{s}$. The figures clearly show that as interaction strength increases (by increasing either $\sigma_0/m_\chi$ or $\eta$), the spur is increasingly suppressed. 
    }
    \label{fig:vary_both_spurs}
\end{figure*}

This section explores how the key findings presented in \secref{sec:massloss} and \secref{sec:stream} change with variations to the SIDM model. In particular, we present results from the SIDM1 and SIDM2 suites which test variations to the normalization of the SIDM cross section $\sigma_0$ and variations to the velocity dependence of the cross section,  parametrized by $\eta$.

\figref{fig:vary_both_isolations} summarizes the fractional mass loss of stars~(dashed lines) and dark matter~(solid lines) as the satellite evolves from infall to first apocenter. The left panel corresponds to the SIDM1 suite (variations in $\sigma_0/m_\chi$ with $\eta = \infty$) and the right panel corresponds to the SIDM2 suite (variations to $\eta$ with $\sigma_0/m_\chi = 30$~cm$^2$/g). For comparison, the results for CDM are indicated by the left-most vertical brackets on each panel. The bracket indicates the full range of values obtained  by repeating the CDM best-fit simulation 11 times, changing only the initial random seed. This quantifies some of the systematic uncertainty from the choice of random initialization seed used in the simulations. Similar uncertainties should be present in the SIDM runs, but have not been verified in this study due to the large number of simulations that would be required to vary those seeds as well.

The different colors in the figure indicate how the results change when only satellite--satellite~(green), satellite--host~(pink), or both types of self interactions~(blue) are active. The different amounts of fractional mass loss for each of these alternatives is useful for evaluating which SIDM processes dominate the satellite's evolution. For example, satellite--satellite interactions are those that are responsible for heat flow within the inner region of the satellite resulting in thermalization of the core and potentially in eventual core collapse. Changes to the satellite's dark matter density profile induced by this heat flow could affect mass loss via tidal stripping. On the other hand, satellite--host interactions are responsible for ram-pressure evaporation. Additional secondary effects could also be present if, for example, the SIDM interactions induce mass loss that affects dynamical friction and the satellite's orbit, which also changes tidal stripping properties. However, such secondary effects would be expected to produce visible changes in the orbits, which we do not observe (see \figref{fig:vary_orbits} and discussion in \appref{appx:orbits}).

Comparing the green curves in each panel to the CDM results for both dark matter and stars, the fractional mass loss is almost unchanged for all SIDM implementations. This indicates that satellite--satellite interactions for these parameter points do not substantially change the tidal stripping properties of a satellite on a Sgr-like orbit. One explanation could be that the baryonic properties of the Sgr dwarf in our simulations prevent substantial coring of the satellite's dark matter density profile~\citep{Ren:2018jpt}. Indeed, we see little evidence for core formation or collapse before the satellite is significantly disrupted by its first pericentric passage (see \appref{appx:profiles}).

Next, consider the pink curves in the left panel of \figref{fig:vary_both_isolations}. For velocity-independent cross sections, the ram-pressure evaporation rate in \eqref{eq:rp} is simple to understand since $\tilde{\sigma}$ is constant in the averaging of $\langle \chi_e \tilde{\sigma} v_{\rm rel} \rangle$. Therefore, for equivalent orbits (as is approximately the case here) the rate per bound satellite dark matter particle is just proportional to $\sigma_0$ and occurs mostly at pericenter when velocities and host particle densities are largest. However, for very large evaporation rates for which most of the dark matter is removed within the first pericenter, the total fractional mass loss of the satellite should slow down with increasing $\sigma_0$, as can be seen from the solution to~\eqref{eq:rp_2} for constant $\Gamma_{\rm rp}$ from which fractional mass loss goes like $(1-\exp{[-\Gamma_{\rm rp}\Delta t]})$.\footnote{$\Delta t$ can be thought of as the time it takes for a single pericentric passage.} In this case, satellite--host interactions become less prevalent as more particles are removed. This is precisely the behavior seen in the solid pink curve.

The pink curve in the right panel of the figure shows the change in fractional mass loss for a velocity-dependent cross section with $\sigma_0/m_\chi = 30$ cm$^2$/g while increasing $\eta$. For such a cross section, the averaging in \eqref{eq:rp} includes the cross section suppression at velocities above $\eta$. Since there is a large hierarchy between the orbital velocity of the satellite and its internal velocity dispersion, the typical velocity difference of interacting particles is approximately just the orbital velocity. Therefore, if at any time during the orbit the satellite moves much faster than $\eta$, the ram-pressure evaporation rate rapidly decreases. Since the rate is also proportional to $\rho_{\rm host}(\vec{x})$, the behavior becomes non-trivial and depends on the interplay between host density, satellite velocity, and the values of $\sigma_0$ and $\eta$. In comparison to the velocity-independent case for the same value of $\sigma_0/m_\chi$, the ram-pressure evaporation rate should begin to decrease as soon as $\eta$ becomes of order or smaller than the orbital velocity, particularly at pericenter where host density is highest. Typical orbital velocities of the satellites in our simulation suites are $\sim360$ km/s at pericenter and $\sim60$ km/s at apocenter.

This behavior is apparent in the right panel of Fig.~\ref{fig:vary_both_isolations} where a considerable increase in fractional dark matter mass loss is observed for $\eta$ values beyond $\sim300$ km/s for simulations implemented with satellite--host interactions (blue and pink solid curves). For smaller values of $\eta$, the fractional mass loss more closely matches the CDM result. On the other hand, simulations with only satellite--satellite interactions (green solid curve) closely match the CDM result for all values of $\eta$. The behavior of stellar mass loss follows the dark matter behavior since removal of dark matter also increases the efficiency of stellar tidal stripping.

We note that the satellite--host only implementation appears to systematically produce larger fractional mass loss than the satellite--host and satellite--satellite implementation (the solid pink curve is slightly above the solid blue curve for all velocity-dependent simulations in the figure). However, the size of this effect is relatively small and is within the typical uncertainty arising from random seed initialization found in the CDM simulation. The effect could therefore simply be a result of this uncertainty. To fully test this hypothesis, we would have to run additional simulations with variations of random seed initialization for the SIDM examples. However, since this does not qualitatively change any of our results or conclusions, we defer such tests to future work. 

\figref{fig:vary_both_spurs} shows the morphology of the spur in the resulting stellar stream for the SIDM1 and SIDM2 simulation suites. The results are consistent with the intuition described above; the height of the spur is a proxy for the mass of the satellite at last pericenter. Since both rows correspond to increasing ram-pressure evaporation rates, and therefore decreasing satellite masses, from left to right the spur also becomes less pronounced as expected.

There is an important caveat for the comparison of our velocity-dependent simulation results to more realistic SIDM models. For many realistic SIDM models, the form of~\eqref{eq:SIDM_CS} is an approximation to the viscosity cross section. Using the viscosity cross section is more computationally feasible than performing simulations with the full angular dependence of the cross section and has been shown to be a good approximation for the evolution of SIDM  halos~\citep{tulin_yu}. However, the ram-pressure evaporation rate depends on the angle-integrated cross section $\tilde{\sigma}$, which can scale differently with $v_{\rm rel}$, decreasing more slowly than the viscosity cross section~\citep{Girmohanta:2022dog}. Thus, simulations approximating the SIDM cross section using~\eqref{eq:SIDM_CS} for velocity-dependent cases could underestimate the effects of ram pressure at high velocities.\footnote{Note that~\cite{Zeng:2021ldo} used the momentum transfer cross section, which should result in a similar underestimation of the result.} For those cases, obtaining the correct result for both halo evolution and evaporation effects would require using the full angular dependence of the cross section.

\section{Discussion}
\label{sec:discussion}

Sagittarius is a special system for studying dark matter self interactions because of its small pericenter, making it a sensitive probe of ram-pressure evaporation.  Because SIDM can have a significant effect on the Sgr remnant and stream, it should be possible to use the system to constrain the SIDM cross section. However, changes to the orbit or internal properties of the Sgr dwarf, as well as to the Milky Way potential, can also alter the mass loss history of Sgr, which would alter stream features like the spur. Furthermore, in this work, we have not accounted for the effect of SIDM on the Milky Way itself. In order to constrain the SIDM cross section, we would need to account for all of these effects self-consistently while fitting the Sgr stream with a flexible Milky Way and Large Magellanic Cloud~(LMC) model. We note that other dwarf galaxy streams which pass close to the Milky Way \citep[e.g., Orphan-Chenab, Jhelum, Indus, etc.,][]{Erkal+2019,Shipp+2021,Li+2022} would likely have experienced enhanced mass loss at the SIDM cross sections considered here, which would also alter their morphology. 

In addition to affecting dwarf galaxy streams, this enhanced mass loss will result in dwarfs with low dark matter masses at the present day, which can be inferred from velocity dispersion data~\citep[e.g.,][]{mcconnachie,Simon_2019}. This evaporation will be particularly strong for dwarfs with small pericenters since they are exposed to the large dark matter density of the inner Milky Way. In particular, Tucana~III, B{\"o}otes~III, and Triangulum~II all have pericenters smaller than Sgr~\citep[1, 8, and 12~kpc, respectively,][]{Pace+2022}, which suggests they will experience strong evaporation at the SIDM cross sections explored here. More work is needed on these systems to compare their expected mass loss from tidal stripping and from evaporation with observations. Because these systems are significantly out of equilibrium due to tides from the Milky Way, the velocity dispersion may not accurately reflect the enclosed mass. Instead, this comparison should be done by directly comparing mock observations of simulated systems with data. 

\section{Conclusions}
\label{sec:conclusions}

We performed a suite of simulations to determine how the infall of a Sgr-like galaxy is affected by the choice of dark matter model, focusing specifically on CDM and SIDM. For an isolated Milky Way-like galaxy, we ran $N$-body simulations that tracked the evolution of stellar and dark matter debris from the satellite, studying the properties of the resulting dwarf remnant and stellar stream.  We optimized the initial conditions of the satellite to reproduce, as best as possible, the present-day position and velocity observed for Sgr.  While computationally intensive, this is a critical step to enable principled comparisons between CDM and SIDM: if a dark matter model affects the orbit of the satellite, then a different starting point may be needed to recover its present-day phase-space coordinates.

The key findings of the paper are presented for a velocity-independent cross section of $\sigma/m_\chi = 30$~cm$^2$/g. These are reasonably strong dark matter self interactions, and we assume that they only apply to velocity scales commensurate with the satellite's orbit, as allowed by current bounds~\citep{Silverman:2022bhs, slone+23}. For this benchmark SIDM model:
\begin{itemize}
    \item The most substantial difference caused by self interactions is enhanced mass loss relative to CDM. In the SIDM case, the dwarf remnant has an enclosed stellar mass (within 5~kpc) that is close to that expected from \citet{sgr_vasiliev}, but an order-of-magnitude lower for the enclosed dark matter mass.  The remnant is stellar-dominated at present day, with the ratio of stellar to dark matter mass being $\sim 3$. Reproducing both the Sgr phase-space location and dark matter mass appears to be  impossible for this parameter point.  Even a drastic (and likely unphysical) increase in the initial mass of the progenitor does not improve the resulting mass of the remnant today. 
    \item Implementing a novel approach in which self interactions are only turned on between certain simulation particles, we identified that the primary mass-loss mechanism is ram-pressure evaporation, which arises from scattering events between satellite and halo dark matter.  Self-scattering between pairs of satellite dark matter particles has no significant effect on the mass loss.  
    \item The overall stellar stream morphology is similar between CDM and SIDM, with the stream pericenters and apocenters appearing to be largely similar.
    One of the most distinctive differences in the morphology is that the spur feature is less pronounced for SIDM than CDM. 
    For SIDM where ram-pressure evaporation is substantial, the DM morphology differs substantially from that of CDM. While for CDM a clear DM stream is visible, evaporated SIDM particles tend to fall more towards the center of the host and phase mix with the host particles.
\end{itemize}

To understand how these results generalize to different SIDM models, we explored both velocity-independent and velocity-dependent cross sections. The results generalize as expected. For the velocity-independent case, the rate of mass loss from ram-pressure evaporation scales as $(1-\exp{[-\Gamma_{\rm rp} \Delta t]})$. For the velocity-dependent model, the rate of ram-pressure evaporation starts to increase with $\eta$ (for a constant $\sigma_0/m_\chi$) once this velocity scale is larger than the velocity of the satellite at pericenter. These results point to the possibility that ram-pressure evaporation could be used to either search for signals of SIDM, or constrain SIDM parameter space, in the regime where the velocity suppression scale of the cross section becomes of order the orbital velocity of the satellite.

We make certain assumptions in the simulation modeling to  improve computational speed and enable large scans over initial conditions and dark matter model parameters.  For example,  we do not model the growth of the host halo and we also simplify the stellar distribution of the host and satellite, assuming spherical symmetry. Studies such as \citet{penarrubia+2010,Oria+2022} have shown that the stellar distribution in Sgr can affect certain stream features and these would not be captured by our setup.  We also ignore the effect of the LMC, which has been shown to impact properties of the Sgr stream~\citep{tango}. These assumptions certainly affect how well our simulated Sgr system matches observations in detail.  Additionally, the dark matter particles in our SIDM hosts are not allowed to self interact to keep the potential similar to that of the CDM hosts.  To set  robust constraints on SIDM models using Sgr, one would need to account for effects on the host potential as well, which may also impact the dwarf's orbital evolution.

This work demonstrates that dark matter self interactions can leave a distinctive imprint on the evolution of the Galaxy's most expansive stellar stream. Sgr is an excellent system for studying self interactions because its close pericentric passage accentuates the impact of ram-pressure evaporation. 
This also suggests that other dwarf galaxy streams---as well as other dwarf galaxies---with close passages to the Milky Way can serve as laboratories for fundamental tests of dark matter.     

\vspace{1em}
\noindent Animations and three-dimensional visualizations of our simulations are available at \url{https://connorhainje.com/sgr-sidm-viz}.

\begin{acknowledgments}
The authors would like to acknowledge Michael Blanton, David W.~Hogg, Rodrigo Ibata, Chervin Laporte, Adrian Price-Whelan, Robyn Sanderson, and Tomer Volansky for useful discussions.
CH is supported by the National Science Foundation Graduate Research Fellowship under Grant Number~DGE-2234660.
OS was supported during this work by the NSF (Grant Number~PHY-2210498) and acknowledges support from the Yang Institute for Theoretical Physics. OS also acknowledges support from the Simons Foundation.
ML is supported by the Department of Energy~(DOE) under Award Number DE-SC0007968, the Binational Science Foundation (Grant Number~2022287), and the Simons Investigator in Physics Award.
DE acknowledges the support of the Australian Research Council through project number DP220102254.
The simulations presented in this article were performed on computational resources managed and supported by Princeton Research Computing, a consortium of groups including the Princeton Institute for Computational Science and Engineering~(PICSciE) and the Office of Information Technology's High Performance Computing Center and Visualization Laboratory at Princeton University.
\end{acknowledgments}

\software{%
    This work was made possible by a number of open-source software projects, including
    NumPy~\citep{numpy},
    SciPy~\citep{scipy},
    Astropy~\citep{astropy:2013,astropy:2018,astropy:2022},
    scikit-learn~\citep{sklearn},
    Matplotlib~\citep{matplotlib},
    GIZMO~\citep{gizmo}, and
    GalIC~\citep{galicSoftware}.
}

\bibliography{main}{}
\bibliographystyle{aasjournal}

\appendix

\section{Initial conditions scans}
\label{appx:scans}

\setcounter{equation}{0}
\setcounter{figure}{0} 
\setcounter{table}{0}
\renewcommand{\theequation}{A\arabic{equation}}
\renewcommand{\thefigure}{A\arabic{figure}}
\renewcommand{\thetable}{A\arabic{table}}

\begin{figure}[h]
    \centering
    \includegraphics[width=0.5\linewidth]{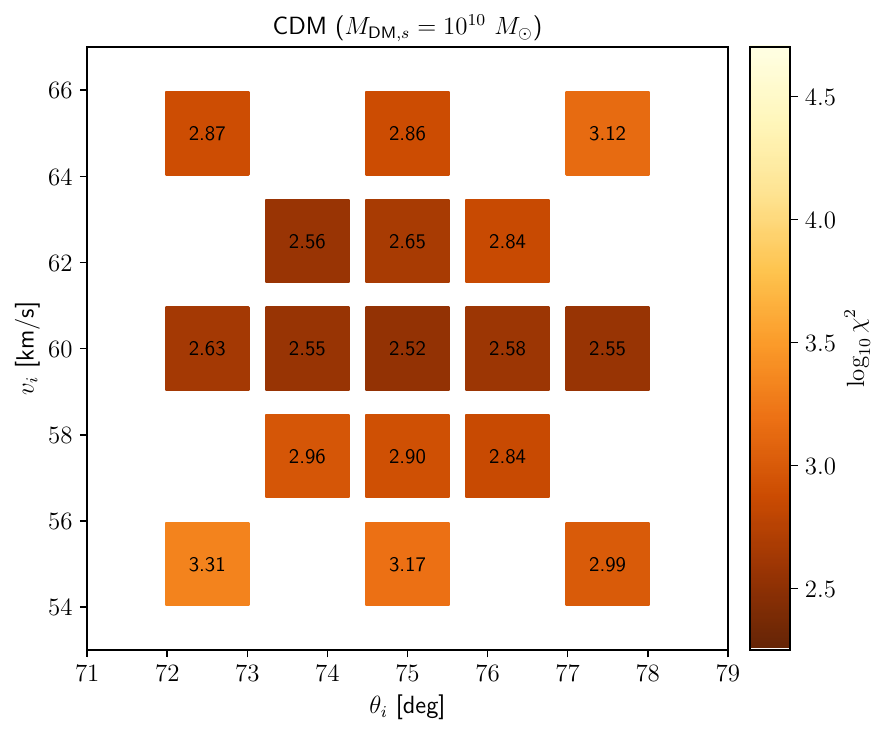}
    \caption{Results of the initial conditions scan for CDM, showing convergence at $v_i = 60 \text{ km/s}$ and $\theta_i = 75^\circ$.}
    \label{fig:scan_cdm}
\end{figure}

\begin{figure}[h]
    \centering
    \includegraphics[width=0.5\linewidth]{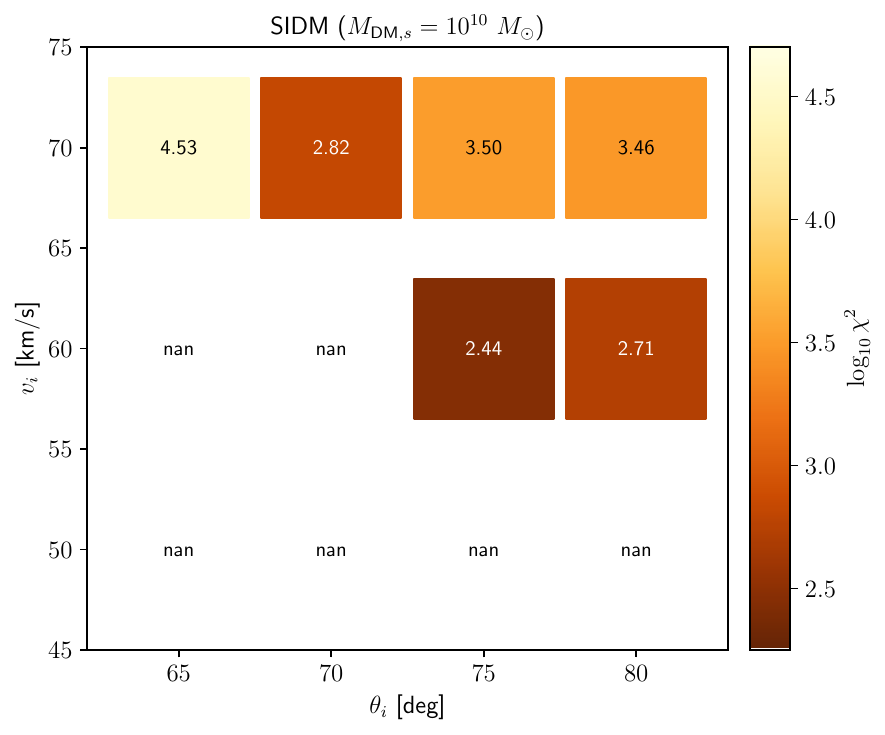}
    \caption{Results of the initial conditions scan for the velocity-independent SIDM model with $\sigma/m_\chi = 30 \text{ cm}^2\text{/g}$ and initial satellite halo mass of $M_{\DM,s} = 10^{10} \ M_\odot$. Points marked \texttt{nan} are simulations where the satellite progenitor was lost before the third pericenter.}
    \label{fig:scan_sidm}
\end{figure}

\begin{figure}[h]
    \centering
    \includegraphics[width=0.5\linewidth]{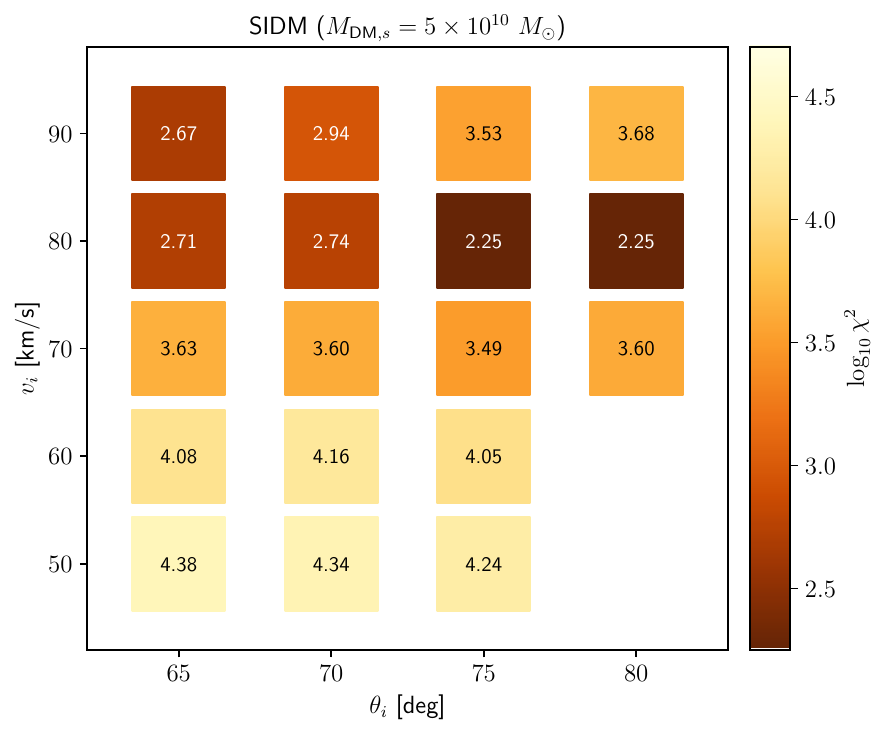}
    \caption{Results of the initial conditions scan for the velocity-independent SIDM model with $\sigma/m_\chi = 30 \text{ cm}^2\text{/g}$ and initial satellite halo mass of $M_{\DM,s} = 5 \times 10^{10} \ M_\odot$. Note that the points $\{ v_i, \theta_i \} = \{ 80 \ \text{km/s}, \, 75^{\circ} \}$ and $\{ 80 \ \text{km/s}, \, 80^{\circ} \}$ have equal $\chi^2$ values to within 0.3\%; we consider the scan converged on $\{ v_i, \theta_i \} = \{ 80 \ \text{km/s}, \, 75^{\circ} \}$.}
    \label{fig:scan_sidm_md}
\end{figure}

\begin{figure}[h]
    \centering
    \includegraphics[width=0.5\linewidth]{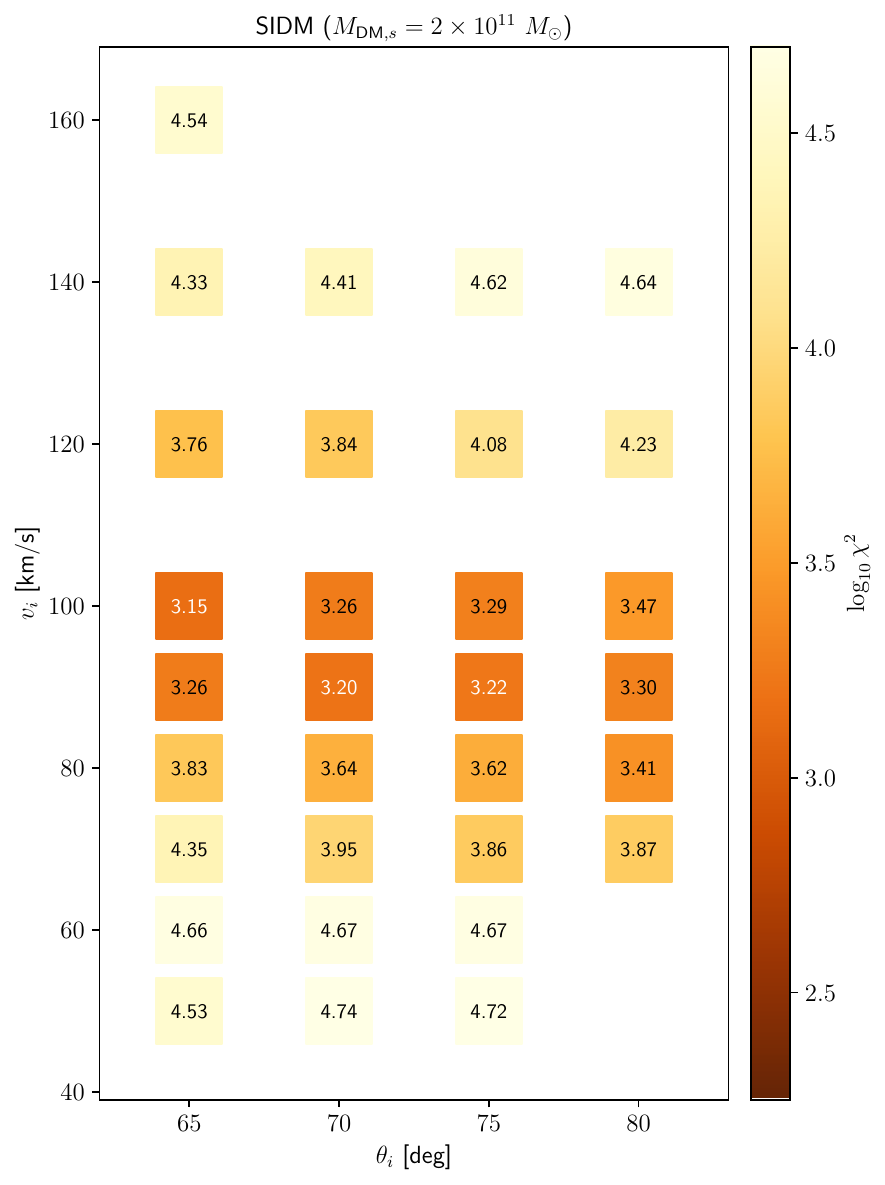}
    \caption{Results of the initial conditions scan for the velocity-independent SIDM model with $\sigma/m_\chi = 30 \text{ cm}^2\text{/g}$ and initial satellite halo mass of $M_{\DM,s} = 2 \times 10^{11} \ M_\odot$. The scan is nearly converged on the point $\{ v_i, \theta_i \} = \{ 100 \ \text{km/s}, \, 65^{\circ} \}$, but we stopped the scan early because the fits were poor.}
    \label{fig:scan_sidm_hv}
\end{figure}

As discussed in \secref{sec:scans}, the CDM~IC and SIDM~IC simulation suites vary over the initial conditions of the Sgr progenitor (namely, the parameters $v_i$ and $\theta_i$) to best match its present-day position and velocity.  This is done by varying $v_i$ and $\theta_i$ over a grid of values to find the point that minimizes the chi-squared value.  Here, we provide the values of $\chi^2$ attained for each simulation as a function of $v_i$ and $\theta_i$. \tabref{tab:scan_results} summarizes the parameters for the best-fit simulations in the CDM~IC and SIDM~IC suites.  Additionally, \figref{fig:scan_cdm} shows the scan results for the CDM~IC simulations, while \figref{fig:scan_sidm}--\ref{fig:scan_sidm_hv} shows the results for the three different progenitor masses used in the SIDM~IC simulations.

In general, we do not achieve values of $\chi^2 \approx 1$. This is due to the small values of $\sigma_x$ and $\sigma_v$ in \eqref{eq:chi2} and the fact that the majority of our simulations do not match the final state to that level of high precision. That being said, the final state is close enough to the target for our purposes, with the Sgr remnant typically falling within a distance of $\mathcal{O}(1)$~kpc and speed of $\mathcal{O}(10)$~kpc from the desired value.

\vspace{0.1in}
\begin{table}[h]
    \noindent
    \centering
    \begin{tabular}{lccccccc}
        \toprule
        Scan
            & $M_{\DM,s}$
            & $v_i$
            & $\theta_i$
            & $\chi^2$
            & $r - r_{\rm tgt}$
            & $v - v_{\rm tgt}$
            & $\theta - \theta_{\rm tgt}$ \\
        
            & [$M_\odot$]
            & [km/s]
            & [$^\circ$]
            & 
            & [kpc]
            & [km/s]
            & [$^\circ$] \\
        \midrule
        CDM~IC
            & $10^{10}$
            & $60$
            & $75$
            & $334$
            & $+0.91$
            & $+15.9$
            & $-0.35$ \\
        SIDM~IC
            & $10^{10}$
            & $60$
            & $75$
            & $273$
            & $+0.61$
            & $+15.7$
            & $-1.42$ \\
        SIDM~IC
            & $5 \times 10^{10}$
            & $80$
            & $75$
            & $179$
            & $+0.80$
            & $+3.8$
            & $-2.59$ \\
        SIDM~IC
            & $2 \times 10^{11}$
            & $100$
            & $65$
            & $1425$
            & $-0.74$
            & $-37.6$
            & $-0.17$ \\
        \bottomrule
    \end{tabular}
    \caption{%
        Properties of the best-fit simulation from each of the CDM~IC and SIDM~IC simulation suites, with the satellite mass $M_{\DM, s}$ specified. $v_i$ and $\theta_i$ are the initial conditions. For the best-fit time of the best-fit simulation, $\chi^2$ gives the value of \eqref{eq:chi2}, and $r - r_{\rm tgt}$, $v - v_{\rm tgt}$, and $\theta - \theta_{\rm tgt}$ give the difference in radius, speed, and angle from the target.
    }
    \label{tab:scan_results}
\end{table}

\clearpage
\section{Orbits for SIDM1 and SIDM2 Suites}
\label{appx:orbits}

\setcounter{equation}{0}
\setcounter{figure}{0} 
\setcounter{table}{0}
\renewcommand{\theequation}{B\arabic{equation}}
\renewcommand{\thefigure}{B\arabic{figure}}
\renewcommand{\thetable}{B\arabic{table}}

\begin{figure*}[h]
    \centering
    \includegraphics[width=0.9\linewidth]{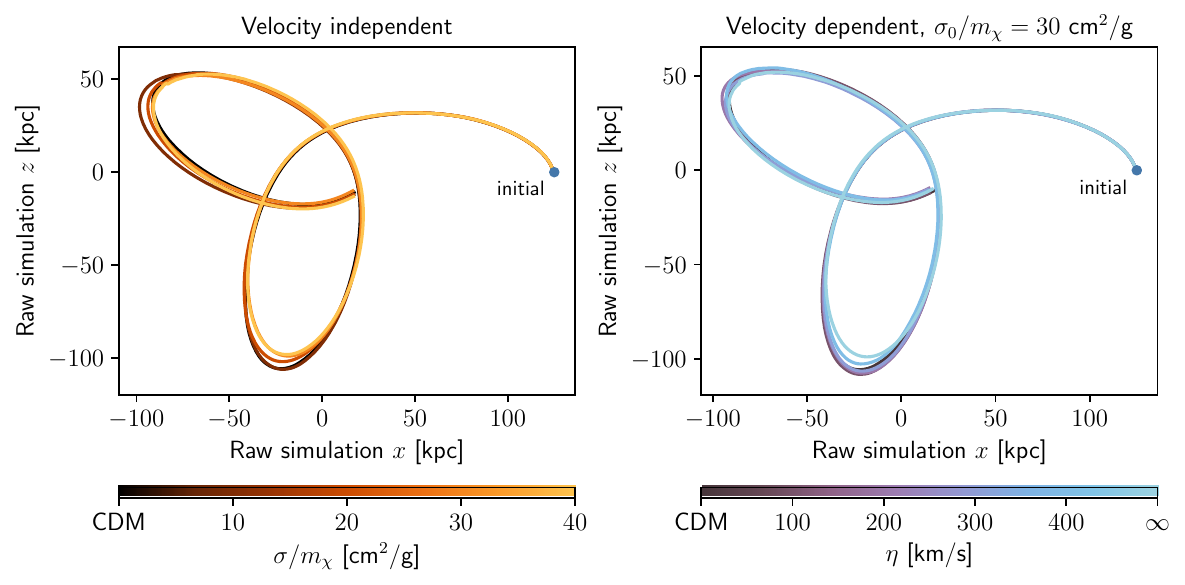}
    \caption{%
        Orbital trajectories of the ``satellite--satellite + satellite--host'' simulations from the SIDM1 and SIDM2 suites, shown in the orbital plane in native simulation coordinates. The left panel shows velocity-independent simulations with cross sections $\sigma/m_\chi = \{ 0, 10, 20, 30, 40 \} \text{ cm}^2/\text{g}$. The right panel shows velocity-dependent simulations with cross section normalization $\sigma_0/m_\chi = 30 \text{ cm}^2/\text{g}$ and velocity scale $\eta = \{ 100, 200, 300, 400 \}$~km/s, as well as the CDM and velocity-independent SIDM ($\eta \to \infty$) simulations. The orbital trajectories are all quite similar, suggesting that the initial conditions adopted for these runs are nearly optimal.
    }
    \label{fig:vary_orbits}
\end{figure*}

The SIDM1 and SIDM2 simulation suites vary the parameters of the SIDM model: namely, the cross section normalization $\sigma_0/m_\chi$ and the velocity scale $\eta$. These simulations are performed using the same initial conditions as those found in the corresponding CDM~IC and SIDM~IC suites.  This assumption is only valid if changing the SIDM model does not strongly affect the satellite's orbit.  To verify this, plots of the orbital trajectories in the native simulation coordinate frame are shown in \figref{fig:vary_orbits} for the velocity-independent~(left) and velocity-dependent~(right) SIDM models. 

\vspace{2em}

\clearpage
\section{Satellite density profiles}
\label{appx:profiles}

\setcounter{equation}{0}
\setcounter{figure}{0} 
\setcounter{table}{0}
\renewcommand{\theequation}{C\arabic{equation}}
\renewcommand{\thefigure}{C\arabic{figure}}
\renewcommand{\thetable}{C\arabic{table}}

\begin{figure*}[h]
    \centering
    \includegraphics[width=0.6\linewidth]{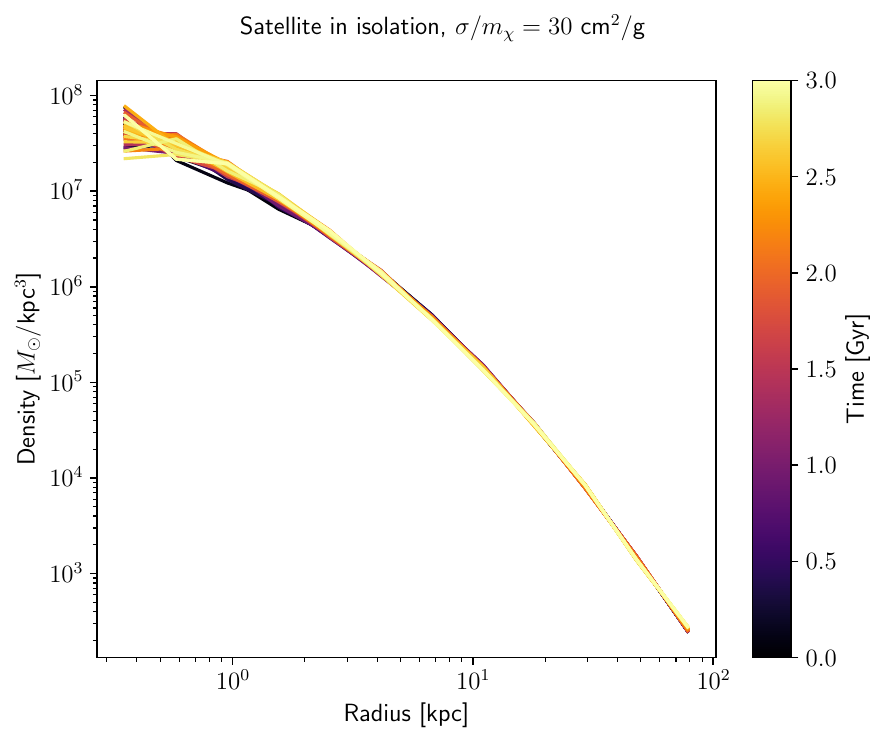}
    \caption{%
        Evolution of the satellite dark matter density profile in a simulation evolving the satellite in isolation with $\sigma / m_\chi = 30$~cm$^2$/g. There is no evidence for the development of a core.
    }
    \label{fig:satellite_isolation}
\end{figure*}

\figref{fig:satellite_isolation} explores whether satellite-satellite interactions affect the initial profile of Sgr at infall. To test this, we run a simulation of the initial satellite \emph{in isolation} with cross section $\sigma/m_\chi = 30 \ \text{cm}^2/\text{g}$. The resulting profile is quite stable with time and does not show any evidence for the development of a core.   This suggests that it is safe to adopt the same initial density profile for the Sgr satellite in both the CDM and SIDM runs, as we do.

\begin{figure*}[h]
    \centering
    Satellite--host only simulations\\
    \includegraphics[width=0.7\linewidth]{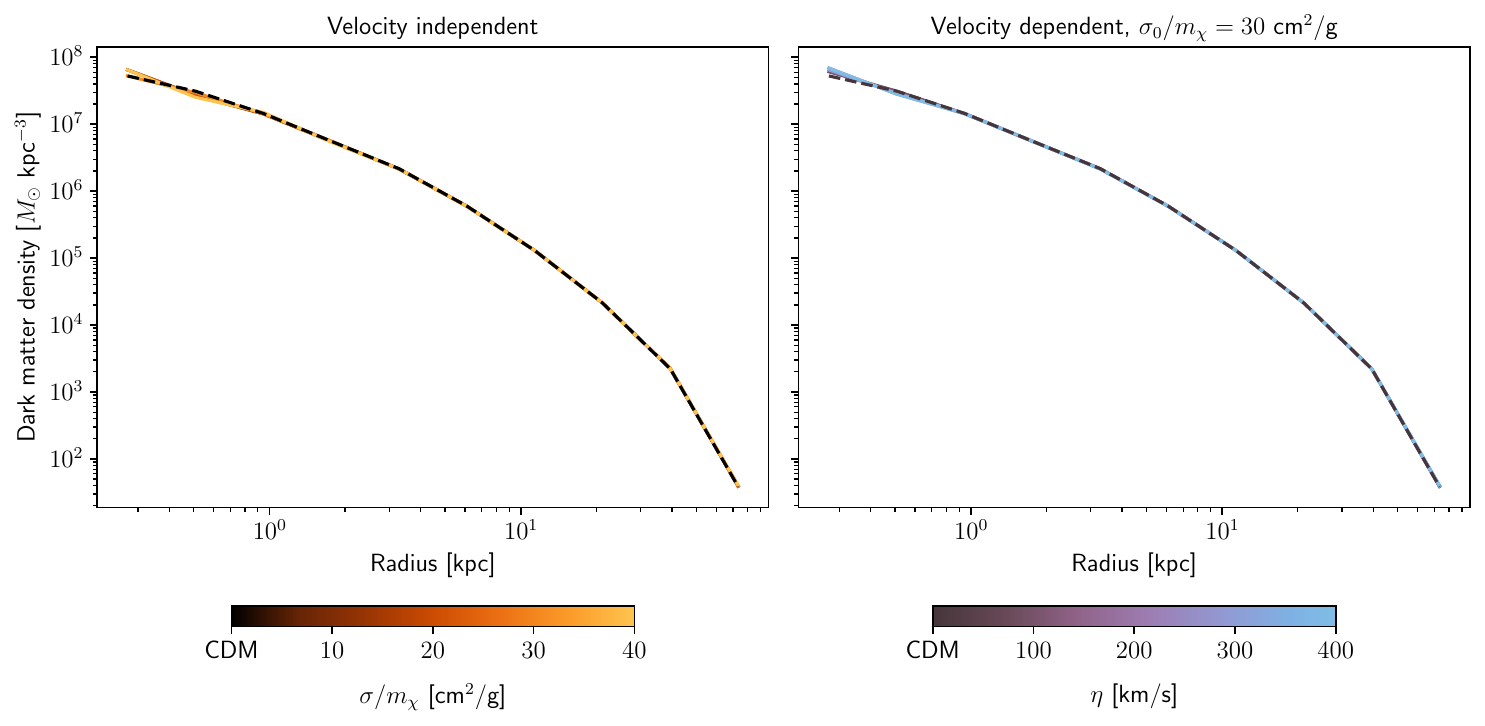}
    \\Satellite--satellite only simulations\\
    \includegraphics[width=0.7\linewidth]{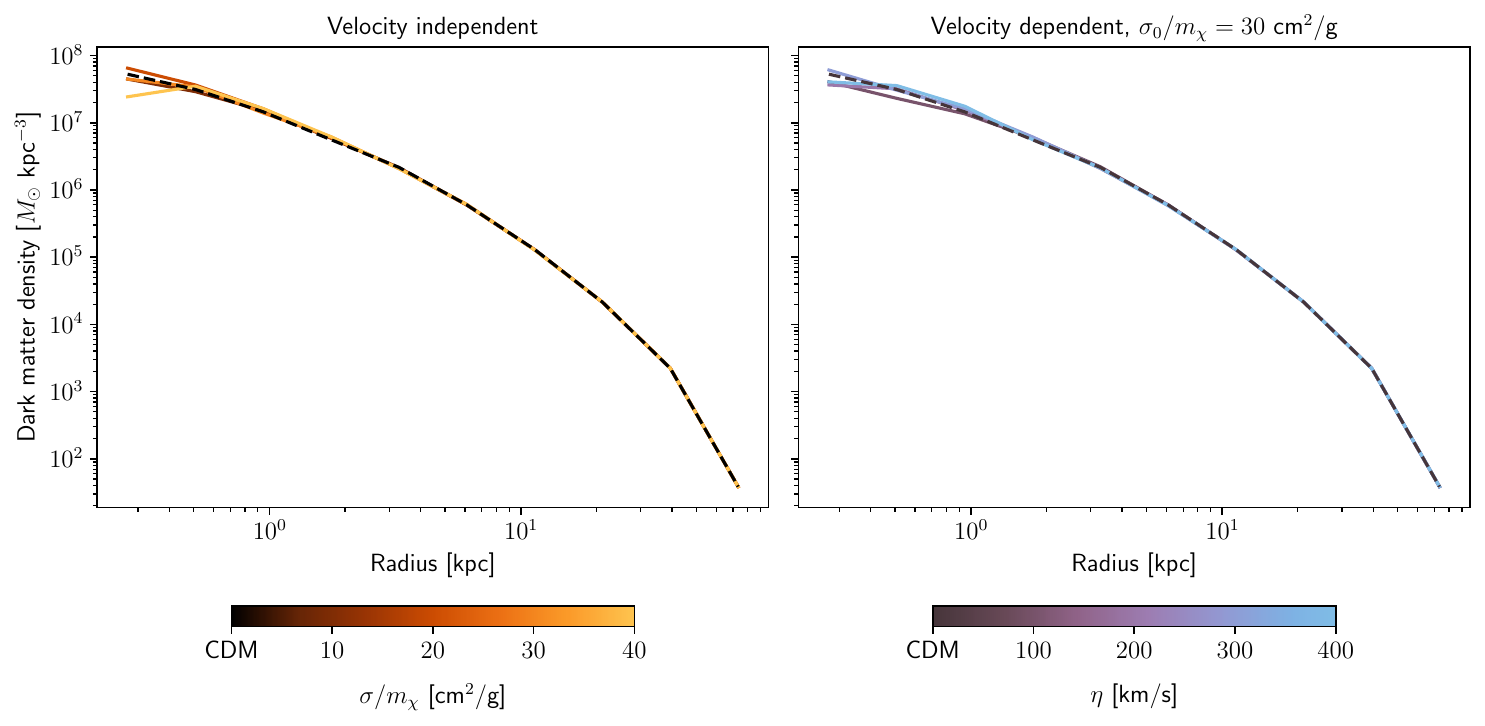}
    \\Satellite--satellite and satellite--host simulations\\
    \includegraphics[width=0.7\linewidth]{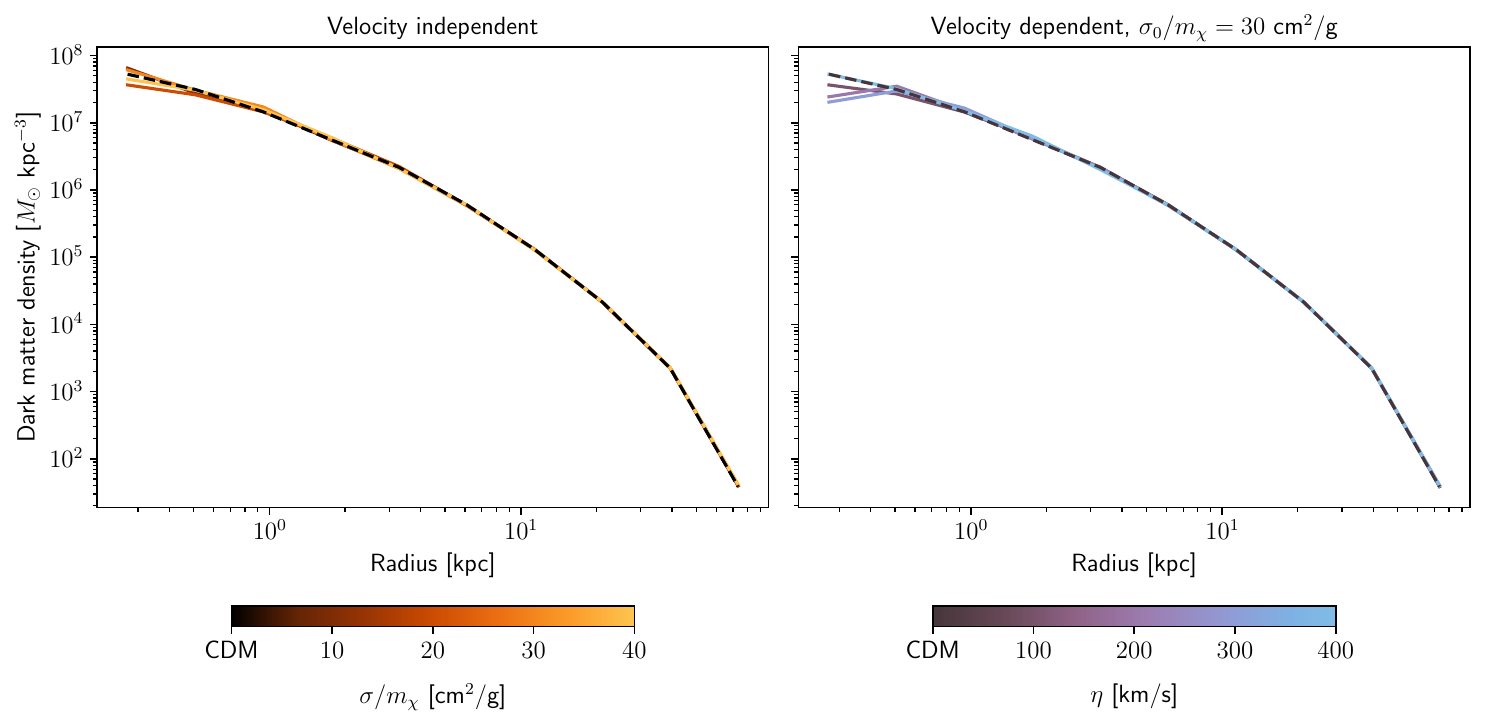}
    \caption{%
        Dark matter density profile of the satellite at 0.5 Gyr, varying SIDM model parameters. Each row corresponds to a different choice of allowed self interactions: the top row shows simulations with only interactions between satellite and host particles (satellite--host only), the middle only between satellite particles (satellite--satellite only), and the bottom has both types of interactions.%
    }
    \label{fig:density_profiles_half}
\end{figure*}

\begin{figure*}[h]
    \centering
    Satellite--host only simulations\\
    \includegraphics[width=0.7\linewidth]{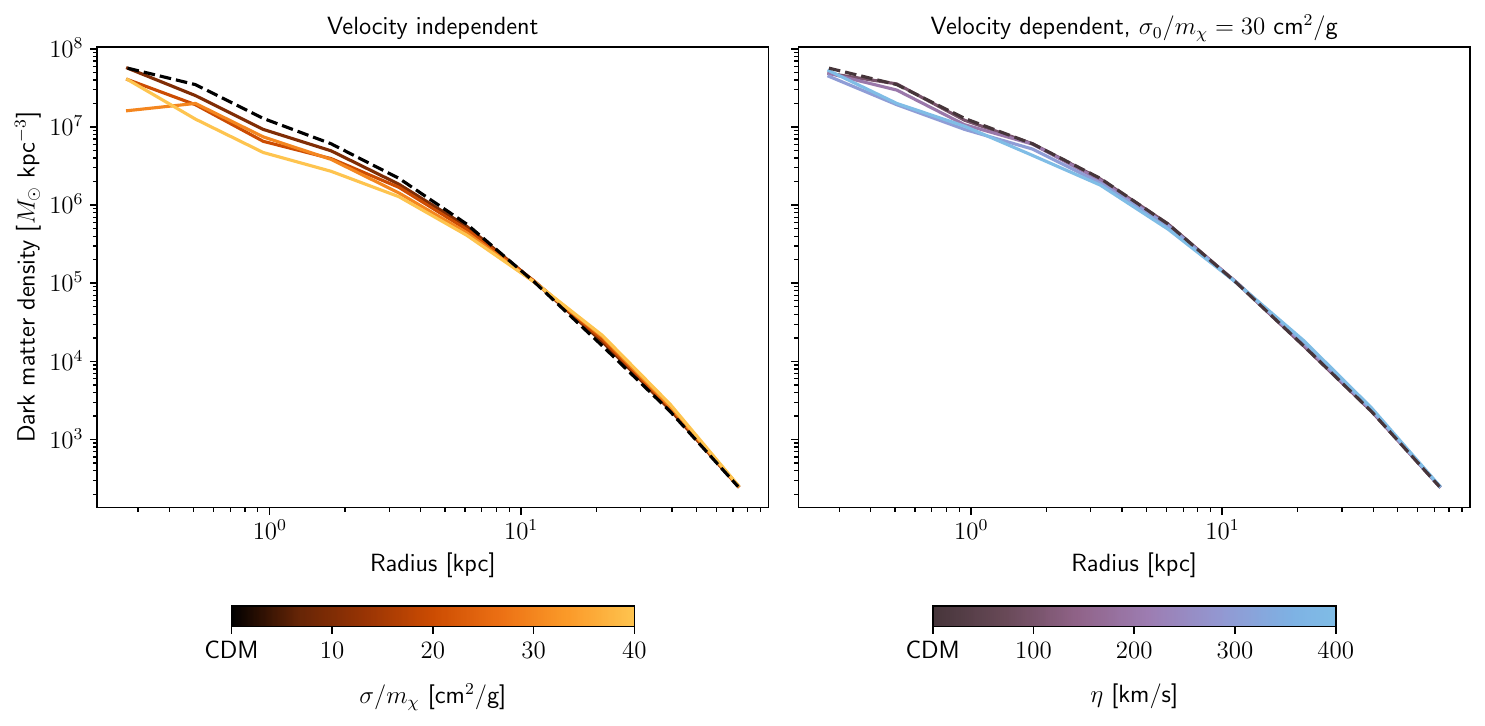}
    \\Satellite--satellite only simulations\\
    \includegraphics[width=0.7\linewidth]{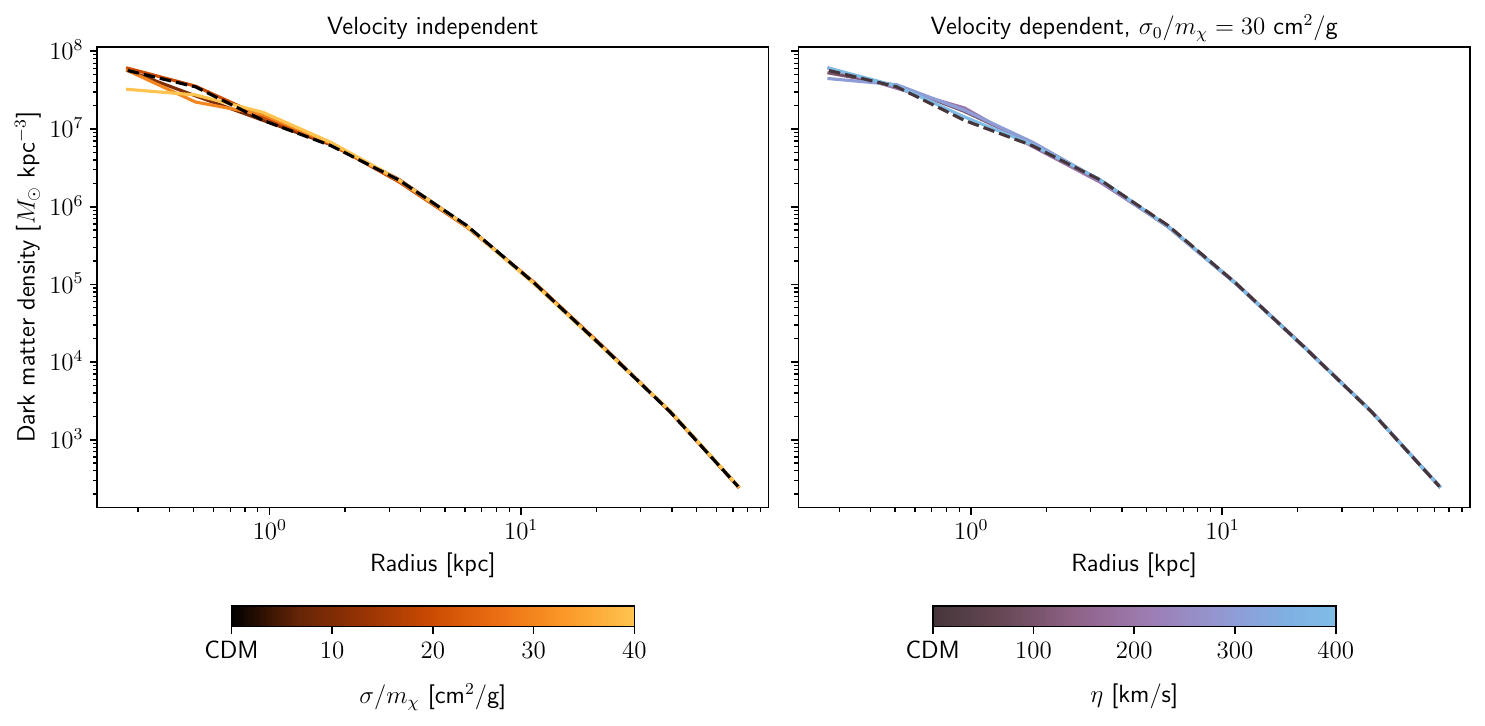}
    \\Satellite--satellite and satellite--host simulations\\
    \includegraphics[width=0.7\linewidth]{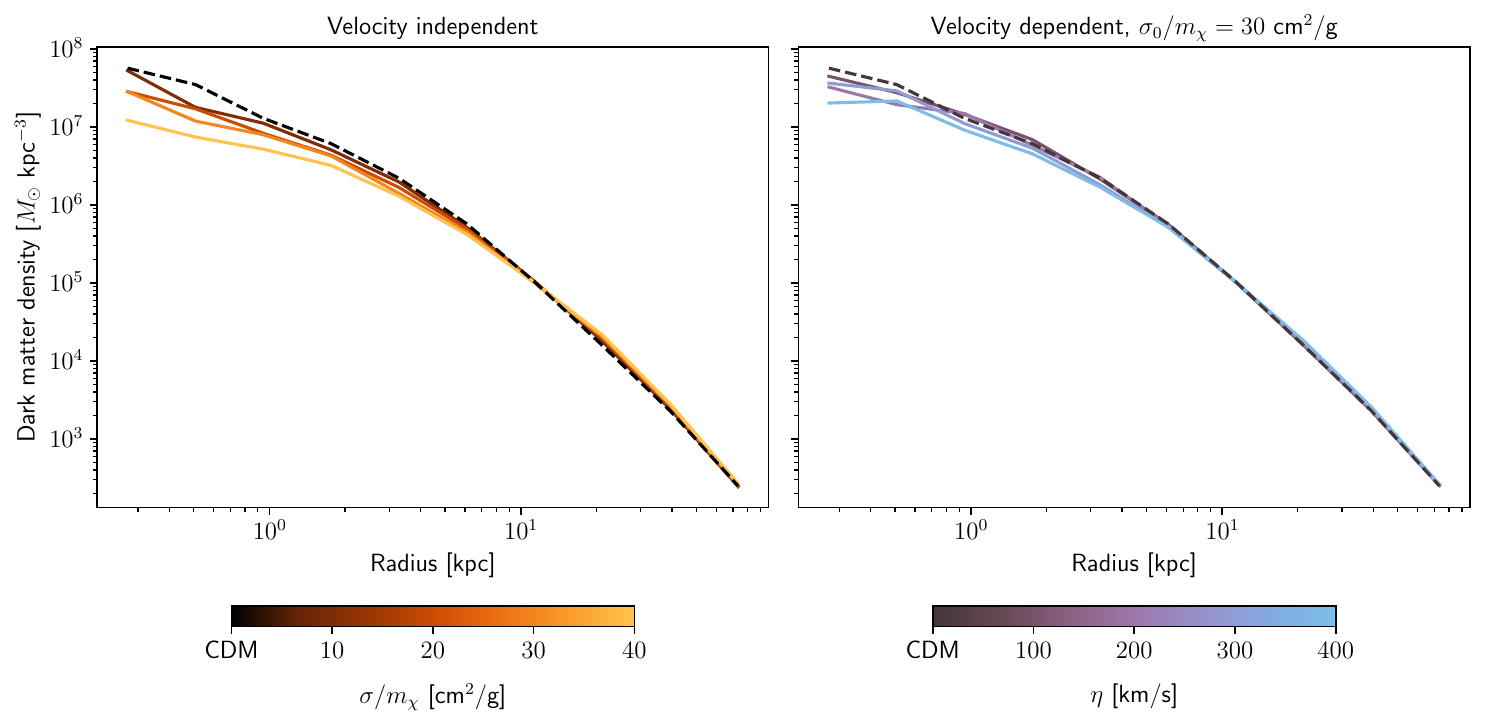}
    \caption{Dark matter density profile of the satellite at 1.0 Gyr, varying SIDM model parameters. Each row corresponds to a different choice of allowed self interactions: the top row shows simulations with only interactions between satellite and host particles (satellite--host only), the middle only between satellite particles (satellite--satellite only), and the bottom has both types of interactions.}
    \label{fig:density_profiles_1}
\end{figure*}

To investigate the degree to which satellite--satellite interactions  reshape the interior of Sgr along its orbit, we present the satellite's dark matter density profile as a function of SIDM model parameters after 0.5~Gyr~(\figref{fig:density_profiles_half}) and 1.0~Gyr~(\figref{fig:density_profiles_1}) of evolution. At both times, we see that the density profile for the dark matter particles that underwent satellite-satellite interactions hardly varies.  This highlights that there is no evidence for core expansion or core collapse of these systems before the satellite reaches first pericenter.

\end{document}